\documentclass[review,11pt]{elsarticle}

\usepackage{epsfig}
\usepackage{amsmath}
\usepackage{amsfonts}
\usepackage{amssymb}
\usepackage{mathdots}
\usepackage{mathdots}
\usepackage{lipsum}
\usepackage[utf8]{inputenc}
\usepackage{enumerate}
\usepackage{amsmath}
\usepackage{amssymb}
\usepackage{centernot}
\usepackage[mathscr]{euscript}
\usepackage{url}
\usepackage{hyphenat}
\usepackage{bm}
\usepackage{verbatim}
\usepackage{accents}
\usepackage{graphicx}
\usepackage{subcaption}
\usepackage{listings}
\usepackage{xcolor}

\usepackage{graphicx}
\usepackage{grffile}
\usepackage{algorithm,algorithmicx}
\usepackage{algpseudocode}
\algnewcommand\algorithmicinput{\textbf{Input: }}
\algnewcommand\algorithmicoutput{\textbf{Output: }}
\usepackage{afterpage}
\usepackage{epstopdf}
\everymath{\displaystyle}

\newcommand\black[1]{\textcolor{black}{#1}}

\newcommand{\Var}{\mathrm{Var}}
\newcommand{\E}{\mathbb{E}}

\journal{Acta Materialia} 

\begin{document}

\begin{frontmatter}

\title{An active learning high-throughput microstructure calibration framework for solving inverse structure-process problems in materials informatics}


\author[add4]{Anh Tran\corref{mycorrespondingauthor}}
\author[add5]{John A. Mitchell}
\author[add4]{Laura Swiler}
\author[add4]{Tim Wildey}
\cortext[mycorrespondingauthor]{Corresponding author: anhtran@sandia.gov}

\address[add4]{Optimization and Uncertainty Quantification Department, Sandia National Laboratories, Albuquerque, NM 87123}
\address[add5]{Computational Multiscale Department, Sandia National Laboratories, Albuquerque, NM 87123}

\begin{abstract}


Determining a process-structure-property relationship is the holy grail of materials science, where both computational prediction in the forward direction and materials design in the inverse direction are essential. 
Problems in materials design are often considered in the context of process-property linkage by bypassing the materials structure, or in the context of structure-property linkage as in microstructure-sensitive design problems. 
However, there is a lack of research effort in studying materials design problems in the context of process-structure linkage, which has a great implication in reverse engineering. 
In this work, given a target microstructure, we propose an active learning high-throughput microstructure calibration framework to derive a set of processing parameters, which can produce an optimal microstructure that is statistically equivalent to the target microstructure. 
The proposed framework is formulated as a noisy multi-objective optimization problem, where each objective function measures a deterministic or statistical difference of the same microstructure descriptor between a candidate microstructure and a target microstructure. 
Furthermore, to significantly reduce the physical waiting wall-time, we enable the high-throughput feature of the \black{microstructure calibration framework} by adopting an asynchronously parallel Bayesian optimization by exploiting high-performance \black{computing} resources. 
\black{Case studies in additive manufacturing and grain growth are} used to demonstrate the \black{applicability} of the proposed framework, where kinetic Monte Carlo (kMC) simulation is used as a forward predictive model, such that for a given target microstructure, the target processing parameters that produced this microstructure are successfully recovered.

\end{abstract}


\end{frontmatter}


\section{Introduction}


Identifying the process-structure-property relationship is the holy grail of materials science, where recent advances in data science and informatics have offered many scalable solutions to accelerate the materials discovery and development \cite{kalidindi2016vision}, which aims to reduce the materials development time by reducing the reliance on physical experimentation \cite{national2011materials,backman2006icme}. 
While substantial efforts have been made in the predictive forward direction, including model calibration for integrated materials computational engineering (ICME) models \cite{chaparro2008material}, high-throughput experiments \cite{salzbrenner2017high}, and high-throughput simulations \cite{ong2019accelerating}, materials design is 
\black{instead equivalent to solving an inverse problem} \cite{sinnott2013material,arroyave2019systems,olson2013genomic}. 
In the forward problem, one asks the predictive question of materials properties and performance characteristics, given a set of materials descriptors related to materials synthesis and processing. 
In the \black{inverse} 
problem, typically, one searches for the optimal and feasible materials descriptors related to materials synthesis and processing, to optimize a set of materials properties and performance characteristics. 
\black{
Notably, while most of the inductive materials design research have been cast directly into the process-structure-property relationship by considering materials properties as objective functions, there is a lack of research \black{involving microstructures} where \black{microstructures are considered as either input or output} in this relationship, particularly with image representation.
However, the approach of bypassing materials structure and seeking direct process-property correlation is not broadly applicable, because it requires a complete description of materials synthesis and processing history, which is often incomplete or not available \cite{kalidindi2016vision}. 
We argue that the materials design problem in inverting process-microstructure relationship is also useful, particularly in reverse materials engineering of a specimen, where only microstructure is available. 
The main challenge in this problem lies in the materials characterization as an effort to obtain a set of useful microstructure descriptors that capture the heterogeneous, anisotropic, and stochastic behaviors of the microstructure. 
}

Deterministic inverse problems, which include inverse problems in materials science contexts, are typically solved using optimization methods. 
In the field of manufacturing, a direct process-property relationship is typically considered, as the goal is to search for the optimal processing parameters which produces single or multiple materials properties. 
In microstructure-sensitive design, one requires an efficient mathematical representation of microstructures and homogenization techniques to establish the structure-property linkage, and searches for the optimal microstructures in the microstructure space that correspond with optimal materials performance \cite{adams2012microstructure,fullwood2010microstructure}. 
We provide a brief literature review on forward prediction of process-structure linkage, inverse problems in process-structure-property relationship, and microstructure-sensitive design problems in Section \ref{sec:LiteratureReview}. 

In our approach, the inverse process-structure problem is formulated as a noisy multi-objective optimization problem, where each objective function is associated with a microstructure descriptor. 
In particular, the objective function measures the deterministic or statistical difference of the same microstructure descriptor applied on two different microstructure images. 
First, a set of microstructure descriptors, which can be deterministic or statistical microstructure descriptors, are used to describe the heterogeneous and anisotropic microstructure. 
Second, some statistical metrics are applied on the probability density functions of these microstructure descriptors to measure the difference between two microstructures. 
This approach results in a set of objective functions, where each objective function corresponds to a microstructure descriptor. 
Multiple objective functions are then \black{scalarized} into a single-objective function, where an asynchronously parallel Bayesian optimization framework is deployed to solve efficiently on a high-performance computing platform. 
To the best of the authors' knowledge, it is the first attempt to solve for the inverse problem in process-structure linkage, where the microstructures are represented as images, which is the most common representation for microstructure.


In the rest of this paper, Section \ref{sec:LiteratureReview} discusses the background on the predictive forward problem in process-structure linkage (Section \ref{subsec:LitRevPSlink}), as well as a brief literature review on inverse problems in process-structure-property linkages (Section \ref{subsec:InvProbPS}), and microstructure-sensitive design (Section \ref{subsec:msSensitiveDesign}). 
Section \ref{sec:Methodology} describes the microstructure calibration methodology, which is then formulated as a noisy multi-objective optimization problem. 
To that end, a parallel Bayesian optimization framework is utilized to efficiently minimize the difference between the target and candidate microstructures, resulting in the optimal microstructure and a set of processing parameters that can yield a similar and statistically equivalent microstructure. 
Section \ref{sec:CaseStudy1} presents \black{the first} case study of kinetic Monte Carlo (kMC) in welding, in which the set of target processing parameters are recovered. 
\black{
Section \ref{sec:CaseStudy2} presents the second case study of kMC in grain growth, where the numerical temperature parameter input is recovered. 
In both sections \ref{sec:CaseStudy1} and \ref{sec:CaseStudy2}, high-dimensional microstructures are considered as outputs. 
}
Section \ref{sec:Discussion} discusses the results and Section \ref{sec:Conclusion} concludes the paper, respectively.

\section{Related works}
\label{sec:LiteratureReview}

In this section, we provide a brief literature review about related works. 
Section \ref{subsec:LitRevPSlink} discusses related work on modeling and simulation of microstructure evolution for a manufacturing process. 
Section \ref{subsec:InvProbPS} discusses inverse problems for the process-structure linkage in computational materials science, which many are related to manufacturing. 
Section \ref{subsec:msSensitiveDesign} reviews previous works in microstructure-sensitive design problems.

\subsection{Process-structure linkage}
\label{subsec:LitRevPSlink}

Numerous ICME modeling and simulation methods have been developed to predict \black{microstructures}, including kMC, finite element, cellular automata, phase field, lattice Boltzmann, and even hybrid methods. 
Rodgers et al. \cite{rodgers2016predicting,rodgers2017monte} implemented a kMC model in SPPARKS to simulate \black{microstructures} for a welding process, to which Li and Soshi \cite{li2019modeling} performed a model analysis on grain morphology with different processing parameters. 
Zhang et al. \cite{zhang2016coupled} proposed a hybrid framework of cellular automata and finite element to predict thermal history and grain morphology in Ti-6Al-4V for direct metal deposition. 
Rai et al. \cite{rai2017simulation} combined cellular automata and lattice Boltzmann to predict microstructure evolution for powder bed based additive manufacturing. 
Archaya et al. \cite{acharya2017prediction}, Chen et al. \cite{ji2018understanding,yu2018phase} employed a phase-field model to study the microstructure prediction and microstructure evolution in laser bed fusion and welding processes. 
Wang et al. \cite{tran2019quantifying,cao2019multi,liu2019mesoscale} combined lattice Boltzmann method with phase-field model to study the dendritic growth in solidification process. 
For a data-driven approach, Kalidindi et al. \cite{yabansu2017extraction,brough2017microstructure,popova2017process} developed a two-point statistics approach, called the Materials Knowledge System framework, to establish the relationship between process and structure linkage by principal component analysis, which is a linear-dimensionality reduction framework.





\subsection{Inverse problems in process-structure-property}
\label{subsec:InvProbPS}


Inverse problems in materials science are typically cast directly into process-property linkage, where materials properties are considered as outputs and manufacturing processes are considered as inputs. 
The linkage is often considered as an unknown black-box function, where optimization methods are utilized to find process inputs which minimize or maximize materials properties. 
\black{This} approach is very common, particularly in the field of manufacturing. 
For example, Jiang et al. \cite{jiang2016optimization} optimized the depth of penetration and bead width in laser welding process without considering microstructure. 
Liang et al. \cite{park2011multi,zhang2015study,wang2018simultaneous} utilized a genetic algorithm and a particle swarm optimization method to various manufacturing problems, where a computational model is considered as objective functional evaluators. 
Pfeifer et al. \cite{pfeifer2018process} utilized the sequential Bayesian optimization framework to search for optimal processing conditions for the fabrication of organic photovoltaic thin films.
Fernandez-Zelaia and Melkote \cite{fernandez2019process} also relied on multi-output Gaussian process \black{model} to build a surrogate model between process and property. 
Forsmark et al. \cite{forsmark2015using} bypassed the traditional approach and established directly the process-property linkage for thin-walled high-pressure die castings of the magnesium alloy AM60. 
Nath et al. \cite{nath2019uncertainty} applied Gaussian process \black{model} and global sensitivity analysis to quantify uncertainty in microstructure grain morphology based on the cellular automata simulations of laser direct metal deposition. 
Ling et al. \cite{ling2017high} proposed a sequential learning approach, where the surrogate model is Random Forests but materials properties are considered as outputs. 
The only work that we have found which targets microstructure as inputs is Paul et al. \cite{paul2019microstructure}. In this work, they also employed Random Forests algorithm, as in Ling et al. \cite{ling2017high} to optimize materials properties with respect to microstructure orientation distribution functions. 
Johnson and Arr{\'o}yave \cite{johnson2016inverse} proposed an inverse design framework for process-structure linkage and identify the heat treatment in Ni-rich NiTi shape memory alloys with a desired size distribution of Ni$_4$Ti$_3$ precipitates. 
To the best of the authors' knowledge, we are unaware of any prior work considering microstructure images as outputs in the process-structure linkage, which is the focus of this study. 


\subsection{Microstructure-sensitive design}
\label{subsec:msSensitiveDesign}

\black{Microstructure-sensitive design problems aim to tailor the microstructure in such a way that materials properties and performance characteristics are optimized \cite{adams2012microstructure,fullwood2010microstructure}. }
First, given a microstructure, a mathematical representation, such as two-point statistics, can be devised to describe the microstructure in a low-dimensional space \cite{kalidindi2016vision,arroyave2019systems} to improve the efficiency of the optimization method. 
Next, homogenization techniques are utilized to form a structure-property linkage, followed by the goal-oriented search in the microstructure space to identify optimal microstructures. 
\black{The} crystal plasticity finite element technique is one of the most widely used tools in this field. 
Fast et al. \cite{fast2008application} demonstrated the first successful design case studies to optimize performance of structural components made from polycrystalline metals with hexagonal close-packed crystal lattices. 
Priddy et al. \cite{priddy2017strategies} employed the Materials Knowledge System framework to significantly reduce the computational cost of crystal plasticity finite element methods and efficiently explore the extreme value distribution of the fatigue indicator parameters with various microstructures. 
Paulson et al. \cite{paulson2018data,paulson2019reduced} followed Priddy et al. \cite{priddy2017strategies} in applying the Materials Knowledge System framework in the transition fatigue regime. 
Sundararaghavan and Zabaras \cite{sundararaghavan2006design} proposed a finite element multiscale homogenization model to capture the elasto-viscoplastic behavior of a microstructure and applied gradient-based optimization methods to achieve the optimal microstructure. 
Recently, Li et al. \cite{li2018deep} and Yang et al. \cite{yang2018microstructural} employed the generative adversarial framework, which is a deep learning approach, to model the microstructure with a set of low-dimensional latent variables, and subsequently optimize the structure-property where the latent microstructure variables and the materials properties are the inputs and outputs, respectively.

\section{Methodology}
\label{sec:Methodology}

In this section, the inverse problem of calibrating processing parameters via microstructure is described in an active learning context, where a high-throughput optimization framework can be utilized to accelerate the optimization procedure. 
Section \ref{subsec:ProblemFormulation} defines the problem and provides an overview perspective of the proposed method. 
Section \ref{subsec:MaterialsCharacterization} discusses the microstructure descriptors, the conditional microstructure descriptors using filters, and the necessity of microstructure filters in some applications. 
Section \ref{subsec:DiffMeasures} describes the quantitative measures for deterministic and statistical microstructure descriptors. 
Section \ref{subsec:ObjectiveFunctions} describes the conversion of the formulated problem in the materials science context to the multi-objective optimization problems in the mathematical context. 
Section \ref{subsec:HighThroughputOpt} discusses the possibility of parallelizing the Bayesian optimization method \black{as an active learning approach} to accelerate the optimization procedure in the high-performance computing environment.

\subsection{Problem formulation}
\label{subsec:ProblemFormulation}

\begin{figure}[!htbp]
\centering
\includegraphics[width=\textwidth, keepaspectratio]{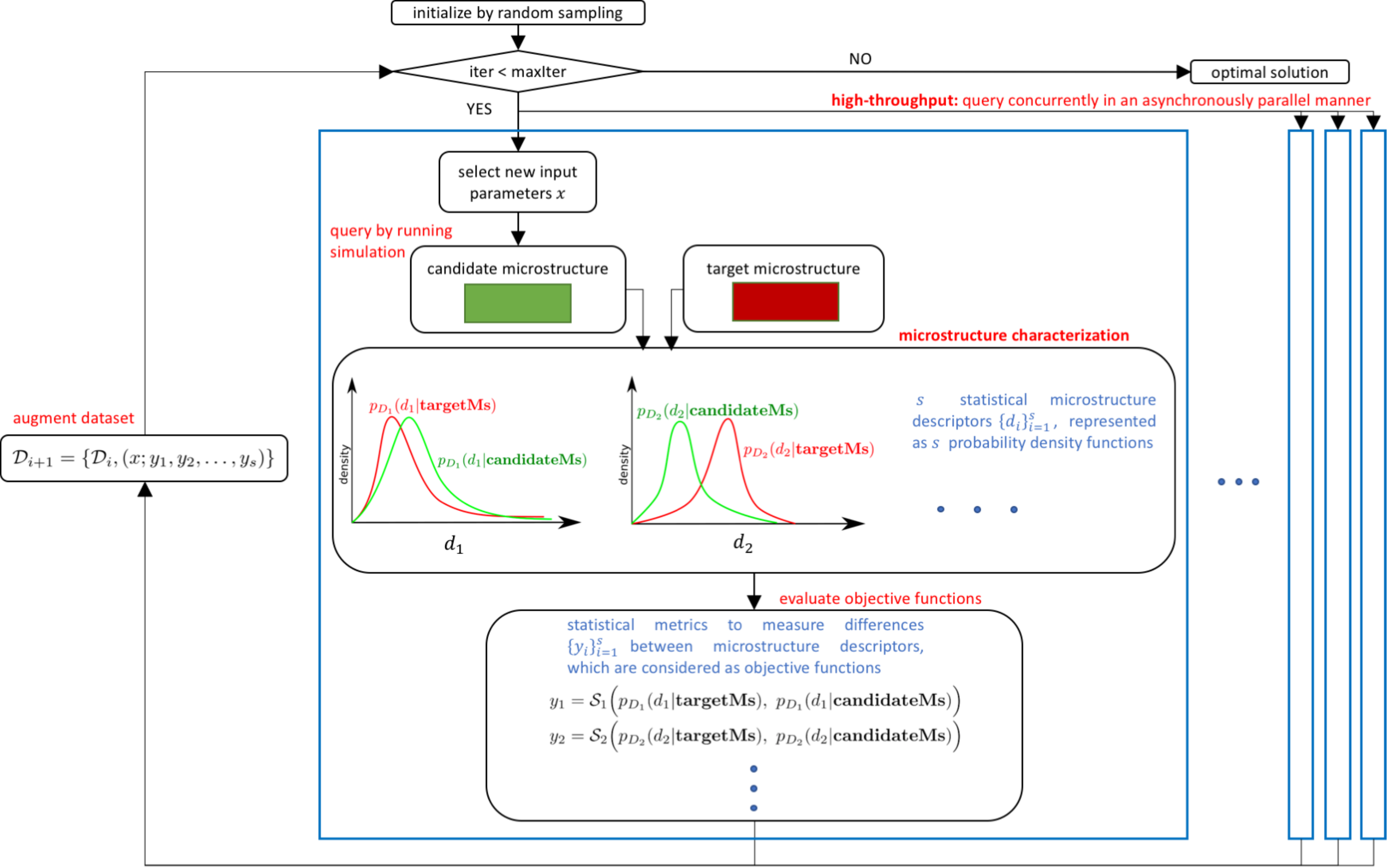}
\caption{An overview perspective of the high-throughput microstructure calibration framework as a multi-objective optimization problem. 
The microstructure descriptors are denoted as $\{d_i\}_{i=1}^s$. 
For statistical microstructure descriptors $d_i$, for some $i$, the microstructure descriptor for $d_i$ are represented as a probability density function $p_{D_i}(d_i)$. 
The microstructure descriptors characterizing the target microstructure and the candidate microstructure are denoted as $p_{D_i}(d_i| \textbf{targetMs})$ and $p_{D_i}(d_i| \textbf{candidateMs})$, respectively. 
For each microstructure descriptor $d_i$, the difference between $d_i$ on the target microstructure and $d_i$ on the candidate is measured through some mathematical or statistical metric $\mathcal{S}_i$, depending on whether $d_i$ is deterministic or statistical. 
The objective functions, denoted as $y_i$, are the difference measured through $\mathcal{S}_i$. 
In particular, if $d_i$ a statistical microstructure descriptor, then $y_i = \mathcal{S}_i \Big( p_{D_i} (d_i | \textbf{targetMs}), \  p_{D_i} (d_i | \textbf{candidateMs}) \Big) $
We assume that there is a one-to-one mapping between the descriptor $d_i$ and $y_i$. 
The \black{microstructure calibration framework} can be generalized to a high-throughput computational materials characterization, where the input $\bm{x}$ can be queried through the prediction model concurrently in an asynchronously parallel manner. 
Readers are referred to the online version for color visualization.}
\label{fig:overviewSchematic}
\end{figure}

Figure \ref{fig:overviewSchematic} presents an overview of the methodology. 
We are concerned with the inverse problem of process-structure linkage, where the target microstructure (denoted as $\textbf{targetMs}$) is known beforehand, and a computational process-structure model that is capable of predicting the microstructure is available. 
The model can be either deterministic or stochastic. 
The goal is to identify the processing parameters $\bm{x}$ that produce the target microstructure. 
In the optimization process, the optimization algorithm proposes many candidate microstructures (denoted as $\textbf{candidateMs}$), and compares quantitatively between microstructures in order to eventually propose an optimal microstructure that matches very closely with the target microstructure in the statistical sense. 
In other words, the optimal microstructure should be statistically equivalent to the target microstructure. 

\subsection{Materials characterization by microstructure descriptors}
\label{subsec:MaterialsCharacterization}

Microstructure is intrinsically noisy due to its stochastic polycrystalline nature, particularly in metallic systems. 
Thus, it is necessary to describe the microstructure using statistical descriptors. 
Bostanabad et al. \cite{bostanabad2018computational}, Liu et al. \cite{liu2013computational}, Li \cite{li2014review}, and Bargmann et al. \cite{bargmann2018generation} provided a comprehensive review for computational microstructure reconstruction, generation, and characterization techniques, with statistical and deterministic physics-based microstructure descriptors. 
\black{
The main difference between physics-based and data-driven microstructure descriptors is their interpretability and explainability. 
On one hand, physics-based microstructure descriptors are explainable and associated with certain physical attributes of the microstructure, e.g. volume of fraction, grain size distribution, aspect ratio, etc. The majority of the work done so far in materials science community falls in the spectrum of physics-based microstructure descriptors. 
On another hand, data-driven microstructure descriptors are impossibly hard to explain, but are capable of fully describing the microstructure. These data-driven microstructure descriptors are often associated with the latent variables in unsupervised machine learning approaches, in particular, deep learning approaches such as generative adversarial networks \cite{latief2010continuum,mosser2017reconstruction,mosser2018stochastic}. 
These latent variables are capable of accurately describing the microstructure and are thus considered as data-driven microstructure descriptors, to contrast with the physics-based microstructure descriptors. 
}

To compare between two microstructures quantitatively and measure how far they are away from each, one needs to impose many microstructure descriptors, including both statistical and deterministic. 
The microstructure descriptors must be able to distinguish one microstructure from another, quantitatively. 
Because of the stochastic nature of microstructure, the statistical microstructure descriptors are used more often than the deterministic ones. 
Suppose that there are $s$ microstructure descriptors, denoted as $\{d_i\}_{i=1}^s$, in Figure \ref{fig:overviewSchematic}. 
Given a microstructure, one can collect a population of grains, and subsequently build a statistics on the  grain population with a probability density function representation. 
For example, one can (i) compute the grain area for each grain in the microstructure, (ii) collect all the observations, and (iii) approximate the probability density function of the observational grain areas of the microstructure.

\subsection{Conditional microstructure descriptors}

It may be beneficial and necessary to impose some conditions 
to compare the microstructure through the lens of statistics, particularly in choosing the right statistics that can distinguish one microstructure from another. 
Conventional statistical microstructure descriptors do not always work very well, mostly due to numerical errors during the computation process. 
If not well chosen, it may be impossible to quantitatively compare microstructures, and thus the optimization framework does not work as one expected. 
Here, we propose to ``filter'' the microstructures to eliminate grains or features that are too common and may negatively impact the optimization process. The filter is implemented by imposing some conditions on conventional statistical microstructure descriptors to create conditional statistical microstructure descriptors.

The process to approximate the conditional statistical microstructure descriptors is described as follows, and is summarized in Algorithm \ref{alg:grainEnsemble}. For each grain $\mathcal{G}_j$ in the filtered microstructure, the grain filter is first applied to eliminate grains that are common, by only considering $\mathcal{G}_j$ if the filter is passed, i.e. $f(\mathcal{G}_j) > 0$. 
Second, a microstructure descriptor $d$ is applied on the grain $\mathcal{G}_j$ after it has passed the filter $f(\cdot)$. 
The quantitative descriptors are then collected for all grains to build a population of samples, where kernel density estimation can be applied to reconstruct the probability density function with the optimal bandwidth. 
Thus, for each statistical microstructure descriptor, there is a corresponding probability density function, constructed via the kernel density estimation method. 

\begin{algorithm}
\caption{Approximation of grain-based microstructure descriptors with probability density functions.}
\label{alg:grainEnsemble}
\algorithmicinput microstructure consisting of $n$ grains $\{\mathcal{G}_j\}_{j=1}^n$, a grain-based descriptor $d$ for $\mathcal{G}_j$, a grain-based filter $f$ for $\mathcal{G}_j$

\algorithmicoutput statistical microstructure descriptors of $p_{D}(d)$
\begin{algorithmic}[1]
\State $\mathcal{C}=\emptyset$ \Comment{initialize sample population}

\For{every grain $\{\mathcal{G}_j\}_{j=1}^n$ in the microstructure}

\If{$f(\mathcal{G}_j) > 0$} \Comment{if the grain passes through the filter}
\State $\mathcal{C} \gets \mathcal{C} \cup d(\mathcal{G}_j)$ \Comment{aggregate the population sample}
\EndIf
\EndFor
\State $p_D(d) \approx \text{KDE}(\mathcal{C})$  \Comment{approx. the density using kernel density estimation}
\end{algorithmic}
\end{algorithm}

In general, given a completely heterogeneous and anisotropic microstructure, the conditional statistics of chord-length density on some particular sampling locations and orientations allow one to quantify the microstructure quantitatively, thus enabling quantitative comparison between one microstructure to another.

\subsection{Measures of differences between microstructures via microstructure descriptors}
\label{subsec:DiffMeasures}

Through microstructure descriptors, one can impose distance metrics to measure the difference between microstructure descriptors of the two microstructures, e.g target and sample microstructure. 
If the microstructure descriptor is deterministic, some mathematical metrics, such as $L^p$-norm, can be utilized to measure the difference. 
If the microstructure descriptor is statistical and can be represented as a probability density function, some statistical metrics and divergences, for example, Wasserstein distance and Kullback-Leibler divergence, can be utilized to measure the difference between probability density functions. 
In particular, if the Kullback-Leibler divergence is utilized to measure the statistical difference $\mathcal{S}(\cdot)$, then the statistical difference between microstructures for a specific microstructure is computed as
\footnotesize
\begin{align}\label{eq:KullbackLeiblerDivergence}
S \Big( p_{D} (d | \textbf{candidateMs}), \  p_{D} (d | \textbf{targetMs}) \Big) 
& =  \mathrm{KL} \Big(p_{D} (d | \textbf{targetMs}) \ \lVert  \ p_{D} (d | \textbf{candidateMs} \Big)\nonumber  \\
& = \int p_{D} (d | \textbf{targetMs}) \ \log \left( \frac{p_{D} (d | \textbf{targetMs})}{p_{D} (d | \textbf{candidateMs})} \right) \partial d &  ,
\end{align}
\normalsize
where KL$(\cdot)$ denotes the Kullback-Leibler divergence and $d$ is the microstructure descriptor. 

\subsection{Objective functions in optimization}
\label{subsec:ObjectiveFunctions}

The differences between these microstructure descriptors are then considered as the objective functions, i.e. $ y_i = \mathcal{S}_i \Big( p_{D_i} (d_i | \textbf{targetMs}), \  p_{D_i} (d_i | \textbf{candidateMs}) \Big), \ i=1,\dots,s $ where the goal is to minimize $s$ objectives, $\{y_i\}_{i=1}^s$, simultaneously. 
Even though a high-dimensional probability density function can be formed by combining several microstructure descriptors together, in this framework, we opt to treat them separately by forming a one-to-one map between microstructure descriptor and objective function. 
It is noteworthy that these objective functions are typically noisy due to the stochastic nature of microstructure, particularly in polycrystalline materials systems. 
As a result, the problem of microstructure calibration can be cast into a noisy multi-objective optimization problem, which can be solved numerically and efficiently on high-performance computing platforms.

\subsection{An active learning high-throughput Bayesian optimization framework}
\label{subsec:HighThroughputOpt}

\black{
Active learning is an area of machine learning that studies the optimal strategies for allocating finite resources. 
The main motivation of active learning approaches is to intelligently select informative points at which the forward objective function should be queried. 
Often in practice, the same accuracy of a supervised machine learning model is achieved, while the size of the training dataset is significantly reduced \cite{settles2009active,settles2011theories}. 
In this context, the Bayesian optimization method arises and is widely regarded as an active learning approach, by balancing the trade-offs between exploration and exploitation. 
On one hand, the exploration advocates for sampling at the least unknown region in the input space. 
On another hand, the exploitation advocates for quick convergence in the best-so-far sample in the training dataset. 
Balancing exploration and exploitation is the key element of the Bayesian optimization to avoid being trapped in local optima through exploration and quickly converge to the global optima through exploitation. 
In the Bayesian optimization method, the trade-off between exploration and exploitation is embedded and reconciled in the acquisition function. 
}

In order to further exploit the available computational resources and expedite the optimization process, multiple instances of the simulations are queried concurrently in a high-performance parallel computing environment to minimize the physical waiting time. 
Parallel optimization methods can be utilized to query the input $\bm{x}$ concurrently to accelerate the optimization procedure. 
For meta-heuristic optimization methods, Memetic et al. \cite{memeti2019using} and Mahdavi et al. \cite{mahdavi2015metaheuristics} provide a comprehensive literature reviews of meta-heuristic optimization methods on high-performance and cloud platforms. 

We discuss the high-throughput optimization method based on surrogate models, in particular, Gaussian process \black{model}, in enabling asynchronously parallel queries. 
The Gaussian process \black{model} is a powerful and widely used tool for approximating black-box functions. 
Bayesian optimization is a natural extension of a Gaussian process \black{model} to globally optimize an unknown black-box function, which is assumed to be smooth. 
By implementing an acquisition function to balance between exploitation (i.e., focusing on promising regions) vs. exploration (i.e., exploring uncertain regions), the traditional Bayesian optimization framework can rely on the Gaussian process \black{model}, as a surrogate model, to search for the global optimal point effectively. 
However, the traditional Bayesian optimization is restricted to sequential query, in which only one input $\bm{x}$ can be queried at once. 

This limitation is indeed a burden on high-performance computing platform, where multiple inputs can be theoretically queried at the same time. 
In our previous work, we have extended the traditional Bayesian optimization framework into synchronously batch-sequential parallel Bayesian optimization \cite{tran2019pbo} for constrained optimization problems, called pBO-2GP-3B, where the constraints are generalized to cover a broad spectrum of applications by considering both known and unknown constraints. 
We deployed a further improved implementation of pBO-2GP-3B, called aphBO-2GP-3B \cite{tran2020aphbo}, to asynchronously parallelize on the high-performance computing environment with multiple acquisition functions considered. 
Readers are referred to our previous work \cite{tran2019pbo,tran2019constrained,tran2019sbfbo2cogp,tran2020smfbo2cogp,tran2020weargp,tran2018weargp,tran2018efficient} and others \cite{brochu2010tutorial,shahriari2016taking,frazier2018tutorial,jones1998efficient} for rigorous literature reviews on Gaussian process \black{model} and Bayesian optimization methods and its variants. 

Regarding multi-objective optimization problems, Alexandropoulos et al. \cite{alexandropoulos2019multi} and Li et al. \cite{li2015many} provide a comprehensive literature reviews on solving multi-objective optimization problems, where the goal is to identify the Pareto frontier. 
Multi-objective optimization problems are typically solved by converting a multi-objective problem to a single-objective problem, where the single objective is a weighted sum of multiple objectives \cite{rojas2019survey}, such as weighted Chebyshev scalarization function $y = \max_{i=1,\dots,s} \lambda_i ( y_i(\bm{x}) - z_i^* )$ ($z_i^*$ represents the ideal value for the objective $y_i$), 
augmented Chebyshev scalarization function $y = \max_{i=1,\dots,s} \lambda_i ( y_i(\bm{x}) - z_i^* ) + \rho \sum_{i=1}^s \lambda_i y_i(\bm{x}) $, and the 
weighted sum scalarization function: $y = \sum_{i=1}^s \lambda_i y_i(\bm{x}) $. 


While there is no restriction on the optimization method that can be used to minimize the difference between the candidate and the target microstructures, in this work, we use the aphBO-2GP-3B Bayesian optimization framework, which is an asynchronously parallel constrained multi-acquisition Bayesian optimization algorithm to further improve its efficiency on high-performance computing platform. 
The aphBO-2GP-3B subdivides the computational budget into three batches, supported by two Gaussian process \black{models}. 
One Gaussian process \black{model} is used to model the objective function, whereas another Gaussian process \black{model} is used as a probabilistic binary classifier for hidden constraints. 
The first batch focuses on optimizing the objective functions by sampling at the locations where the acquisition function value of the Bayesian optimization is maximized. 
The second batch focuses on the exploration, by sampling at the locations where the posterior variance is maximized. 
The third batch focuses on the feasibility classification to learn hidden constraints from the optimization problem. 


\section{Case study \#1: kMC simulation for additive manufacturing and Bayesian optimization}
\label{sec:CaseStudy1}

In this section, we demonstrate our proposed framework through a case study that combines an advanced extension of the Bayesian optimization method and a kMC model for additive manufacturing.

\subsection{Kinetic Monte Carlo simulation}
\label{subsec:kMC}




We adopt the kMC simulation using SPPARKS \cite{plimpton2009crossing,plimpton2012spparks}, where the welding model is developed by Rodgers et al. \cite{rodgers2016predicting,rodgers2017monte,rodgers2017simulation} to predict the microstructure in welding. 
For the sake of completeness of the work, we briefly summarize the kMC used as follows. 
The model simulates melting, solidification, and solid-state microstructural evolution of regions surrounding the weld pool. 
Grain evolution is modeled spatially with three different regions, the base metal, the heat-affected zone, and the fusion zone. 
The base metal is unaffected by the weld pool and thus retains its microstructure. 
In the heat-affected zone, the temperature evolves and induces microstructural evolution, but the temperature in this zone does not exceed the melting temperature. 
In the fusion zone, the temperature is higher than the melting temperature, thus material is completely melted. 
The weld pool geometry is modeled using B\'{e}zier curves to capture a wide range of pool shapes. 

\begin{figure}[!htbp]

\begin{subfigure}[b]{0.475\textwidth}
\centering
\includegraphics[width=\textwidth, keepaspectratio]{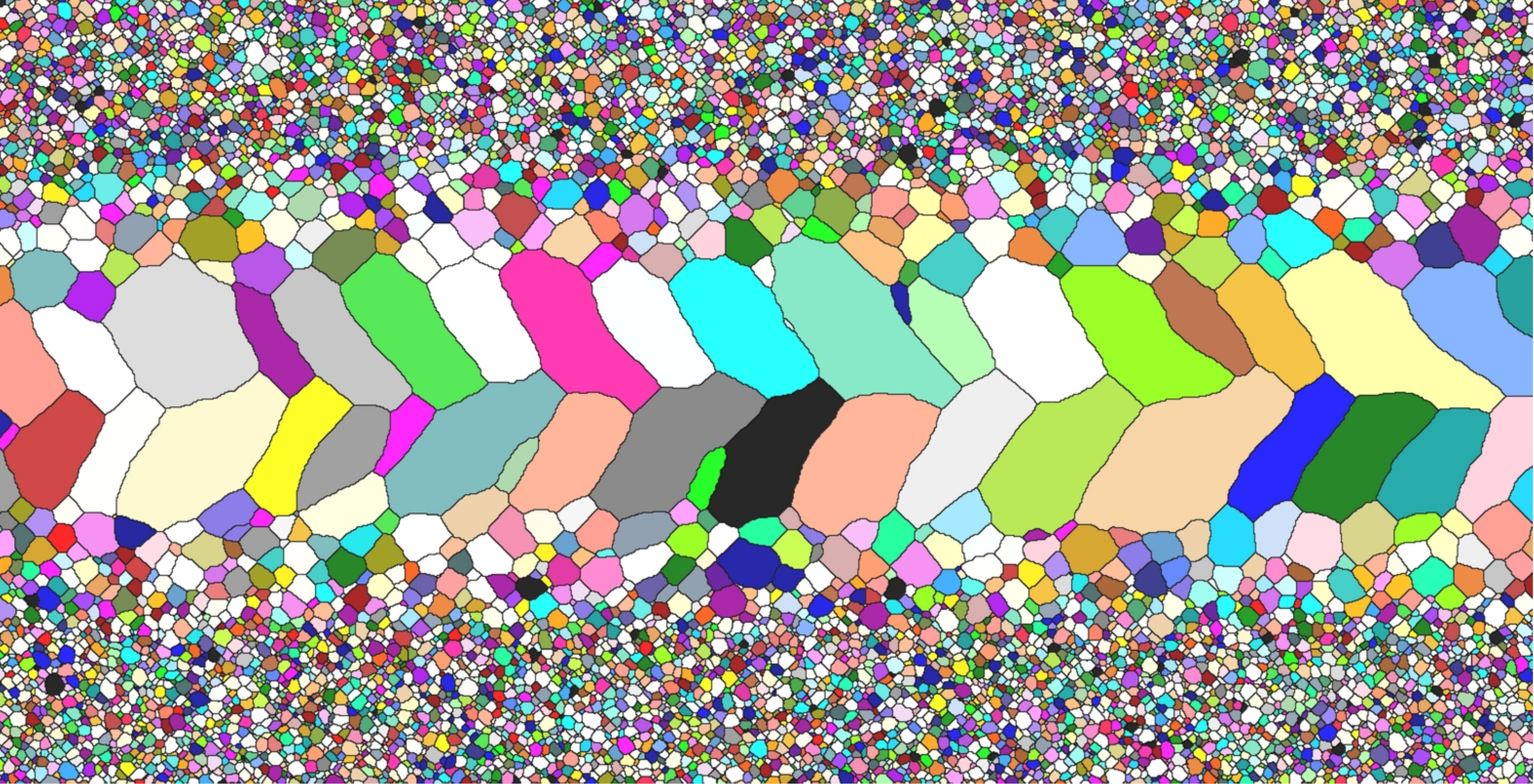}
\caption{Sample 1 (teardrop welding pool shape).}
\label{fig:S}
\end{subfigure}
\hfill
\begin{subfigure}[b]{0.475\textwidth}
\centering
\includegraphics[width=\textwidth, keepaspectratio]{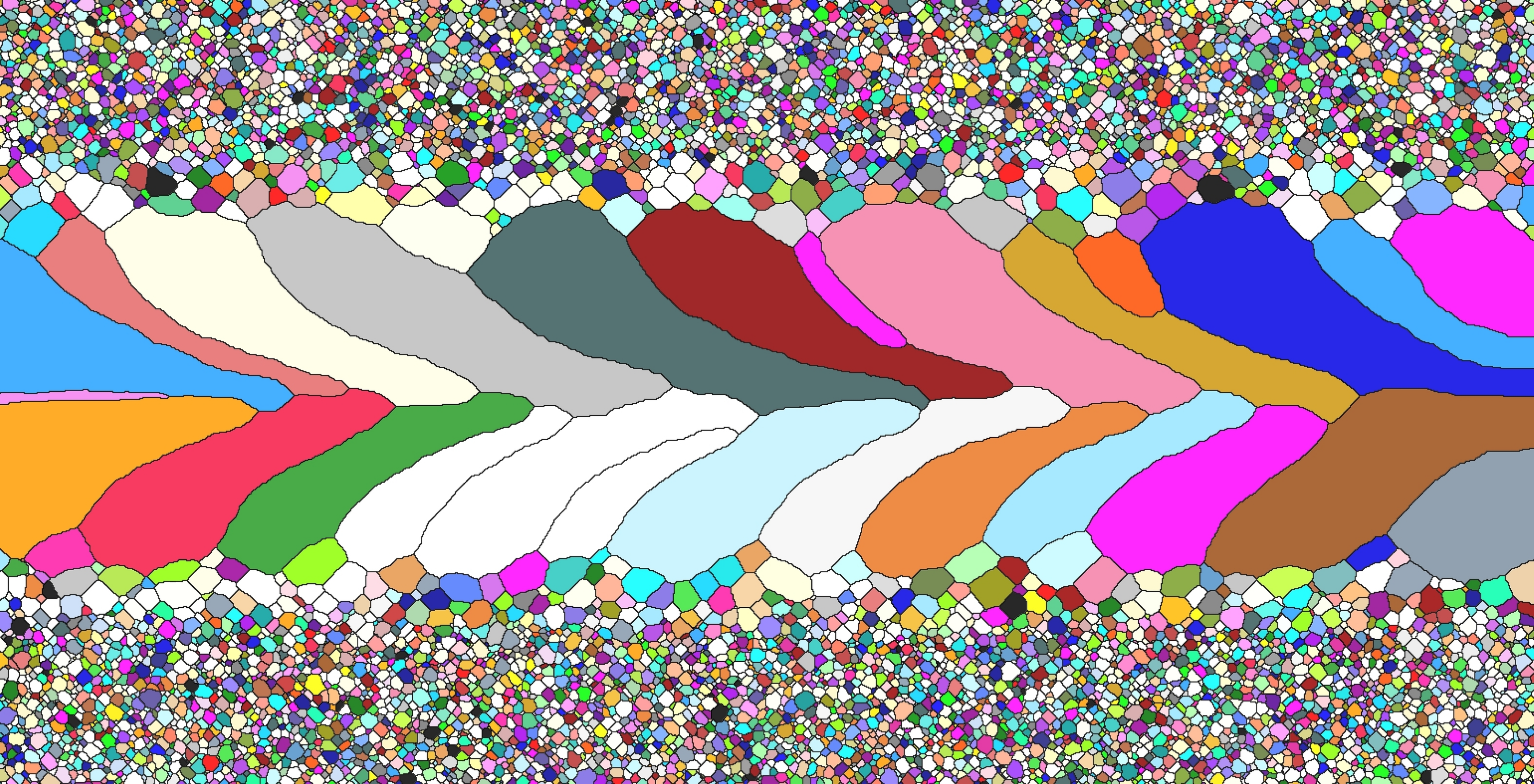}
\caption{Sample 2 (ellipse welding pool shape).}
\label{fig:sample2Ms}
\end{subfigure}

\caption{Two samples of microstructure produced by kMC simulation via SPPARKS.}
\label{fig:samplesOfMs}
\end{figure}

The grain evolution is modeled using a modified Potts Monte Carlo model to capture the process temperature profile as a function of position and time, which is obtained by translating the modeled weld pool and temperature gradient through the simulation domain. 
Grain growth is simulated by the normal curvature-driven grain growth model with grain boundary mobility as a function of temperature. 
In this curvature-driven grain growth model, grain boundary energy is the driving force for grain growth and evolution. 
Grain growth is simulated by minimizing the total grain boundary energy, calculated as the sum of all bonds between neighboring sites belonging to two different grains, multiplied by the bond energy. 
Figure \ref{fig:samplesOfMs} presents two samples of welding microstructure with the kMC model in final state, where each microstructure corresponds to a different set of processing parameters.

The input for the 2D kMC simulation is the velocity, the dimension of the heat-affected zone, and the width of the melting pool. 
For 3D kMC simulations, two more parameters modeling the geometric shape of the welding pool can be considered. 
Table \ref{tab:inputTeardrop} lists the input parameters used to generate the target microstructure in this case study, which closely matches with the work of Rodgers et al. \cite{rodgers2016predicting,rodgers2017monte,rodgers2017simulation}. 
The dimension parameters $W$, $L$, and $H$ are assumed to be constant and thus are fixed during the optimization process.

\begin{table}[!htbp]
\centering
\scriptsize
\begin{tabular}{|c|c|c|c|c|} \hline
\textbf{Input} & \textbf{Physical description} & \textbf{Unit} & \textbf{Lower bound} & \textbf{Upper bound} \\ \hline
$W$              & width of the microstructure   & site          & 805                  & 805                  \\
$L$              & length of the microstructure  & site          & 1575                 & 1575                 \\
$H$              & height of the microstructure  & site          & 1                    & 1                    \\ \hline
$v$              & velocity                      & site/MCS      & 15                   & 30                   \\
$haz$            & heat-affected zone            & site          & 120                  & 200                   \\
$width$          & pool width (teardrop welding pool)                    & site          & 50                  & 250                 \\ 
\hline
\end{tabular}
\normalsize
\caption{Input parameters of the kMC/SPPARKS simulation with teardrop shape welding pool.}
\label{tab:inputTeardrop}

\end{table}

\subsection{Statistical microstructure descriptors as outputs}

In this case study, multiple \black{statistical physics-based microstructure descriptors} are used to quantify the kMC simulated microstructures. 
In the kMC simulation, the microstructure is represented as a ``spin'', which is analogous to the identification number for grains. Sites which correspond to the same ``spin'' belong to the same grain. Using this representation, the grains are segmented, where the grain boundaries is captured. 
The whole microstructure is considered as an ensemble of grains, where the probability density functions are inferred based on the grain observations. 

\begin{table}[!htbp]
\caption{List of microstructure descriptors, their nature, and physical meaning.}
\label{tab:msDescriptors11}
\centering
\scriptsize
\begin{tabular}{|l|l|l|l|l|l|} \hline
\textbf{Notation} & \textbf{Related quantities}         & \textbf{Physical meaning} & \textbf{Type} & \textbf{Nature} & \textbf{Region} \\ \hline
$p_{D_1}(d_1)$                     & $a$: major dim. of best fit ellipse & dist. of $a$          & grain-based & statistic  & global \\
$p_{D_2}(d_2)$                     & $b$: minor dim. of best fit ellipse & dist. of $b$          & grain-based & statistic  & global \\
$p_{D_3}(d_3)$                     & $\theta$: ellipse orientation       & dist. of $\theta$     & grain-based & statistics & global \\
$p_{D_4}(d_4)$                     & grain area                          & dist. of grain area   & grain-based & statistics & global \\
$p_{D_5}(d_5)$                     & chord-length in $x$- direction      & dist. of chord-length & grain-based & statistics & global \\
$p_{D_6}(d_6)$                     & chord-length in $y$- direction      & dist. of chord-length & grain-based & statistics & global \\
$p_{D_7}(d_7)$                     & local chord-length dist., band 0    & dist. of chord-length & grain-based & statistics & local  \\
$p_{D_8}(d_8)$                     & local chord-length dist., band 1    & dist. of chord-length & grain-based & statistics & local  \\
$p_{D_9}(d_9)$                     & local chord-length dist., band 2    & dist. of chord-length & grain-based & statistics & local  \\
$p_{D_{10}}(d_{10})$               & local chord-length dist., band 3    & dist. of chord-length & grain-based & statistics & local  \\
$p_{D_{11}}(d_{11})$               & local chord-length dist., band 4    & dist. of chord-length & grain-based & statistics & local  \\ \hline
\end{tabular}
\normalsize
\end{table}

\begin{figure}[!htbp]
\centering
\begin{subfigure}[b]{0.475\textwidth}
	\centering
	\includegraphics[width=1\textwidth,keepaspectratio]{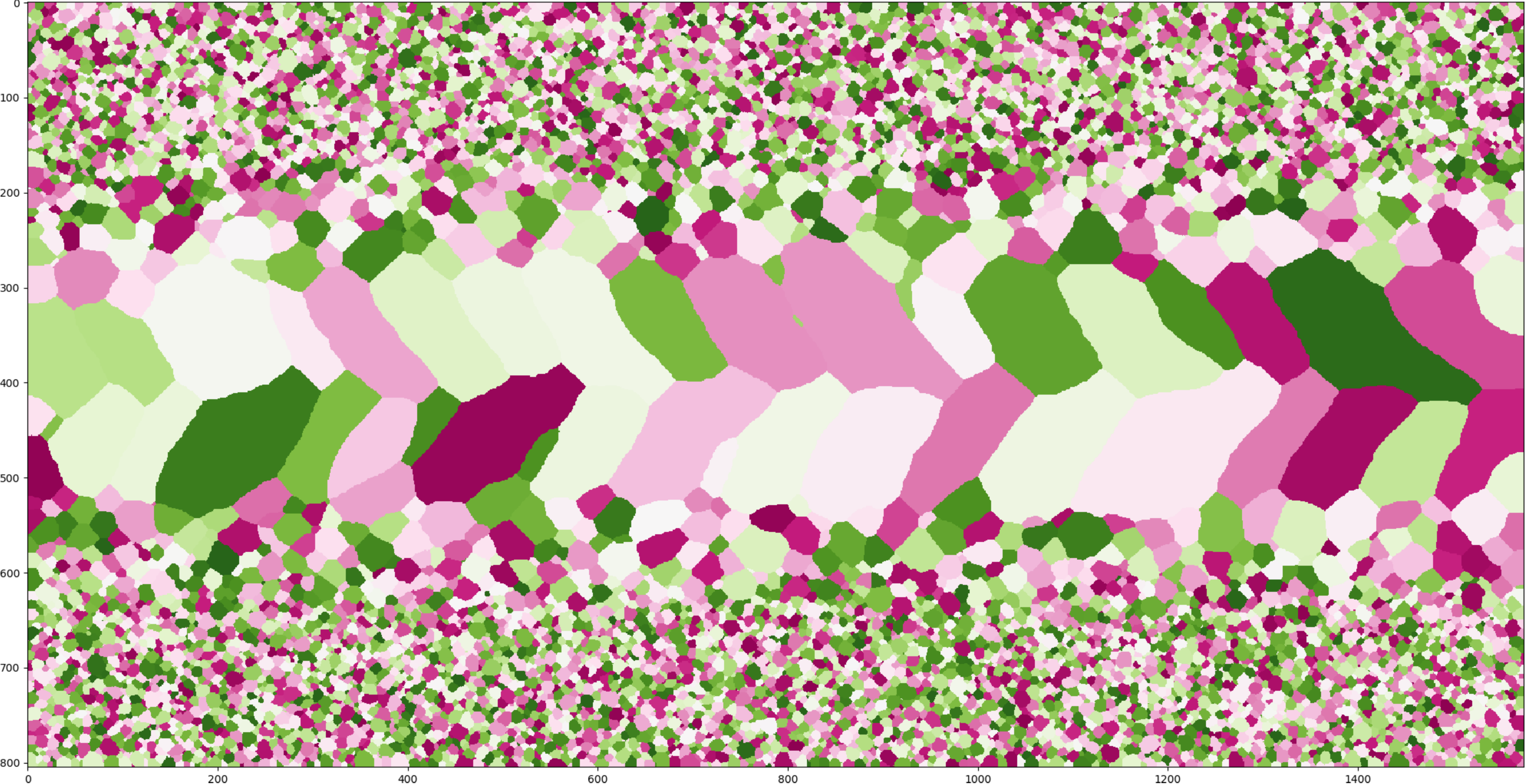}
	\caption{A rasterized ms. representation with dimension.}
	\label{fig:microstructureExample}
\end{subfigure}
	\vfill
\begin{subfigure}[b]{0.475\textwidth}
	\includegraphics[width=1\textwidth,keepaspectratio]{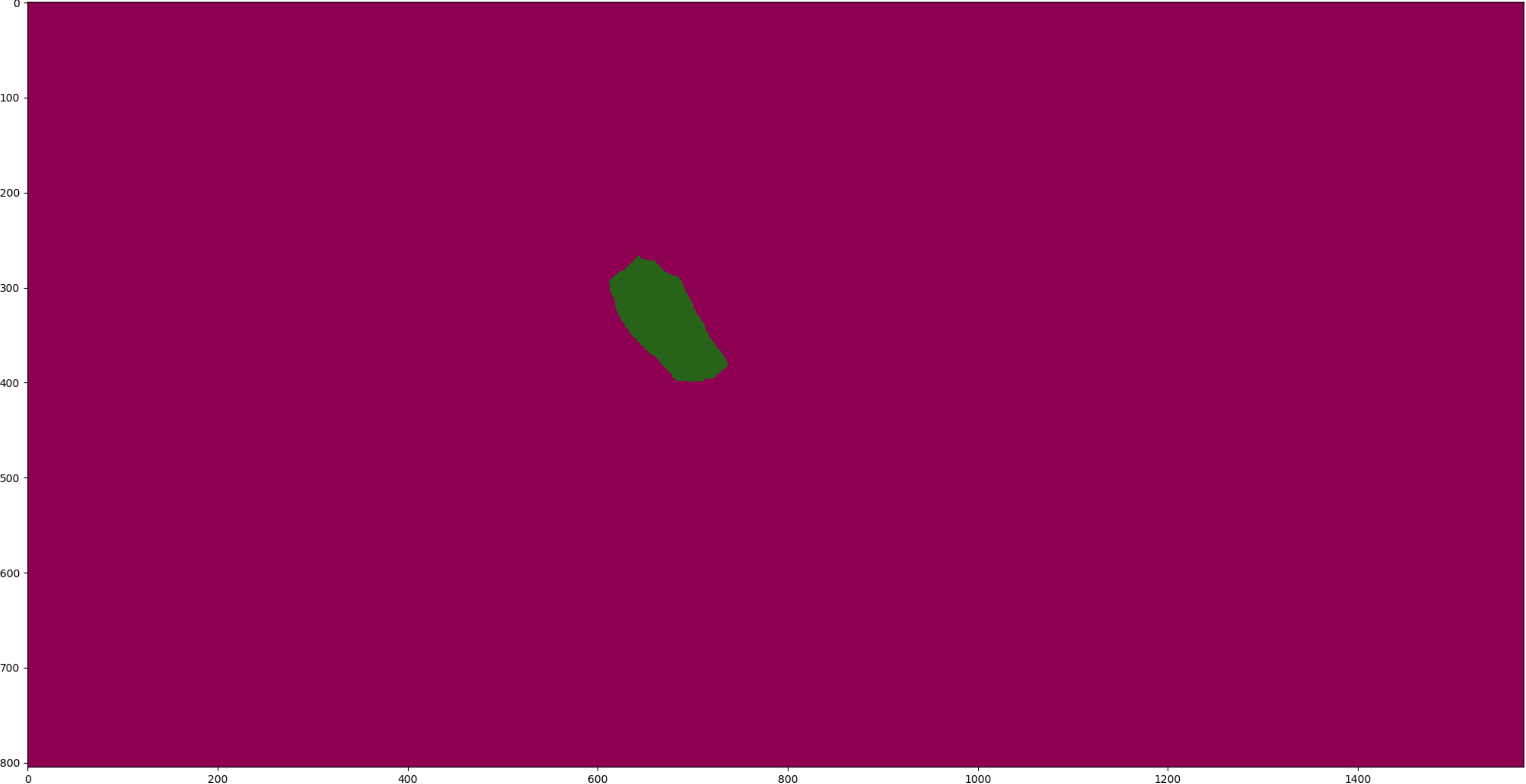}
	\caption{Grain 16166 in the microstructure shown in Figure \ref{fig:microstructureExample}.}
	\label{fig:cropped_uniqueGrain16166}
\end{subfigure}
	\vfill
\begin{subfigure}[b]{0.475\textwidth}
	\includegraphics[width=1\textwidth,keepaspectratio]{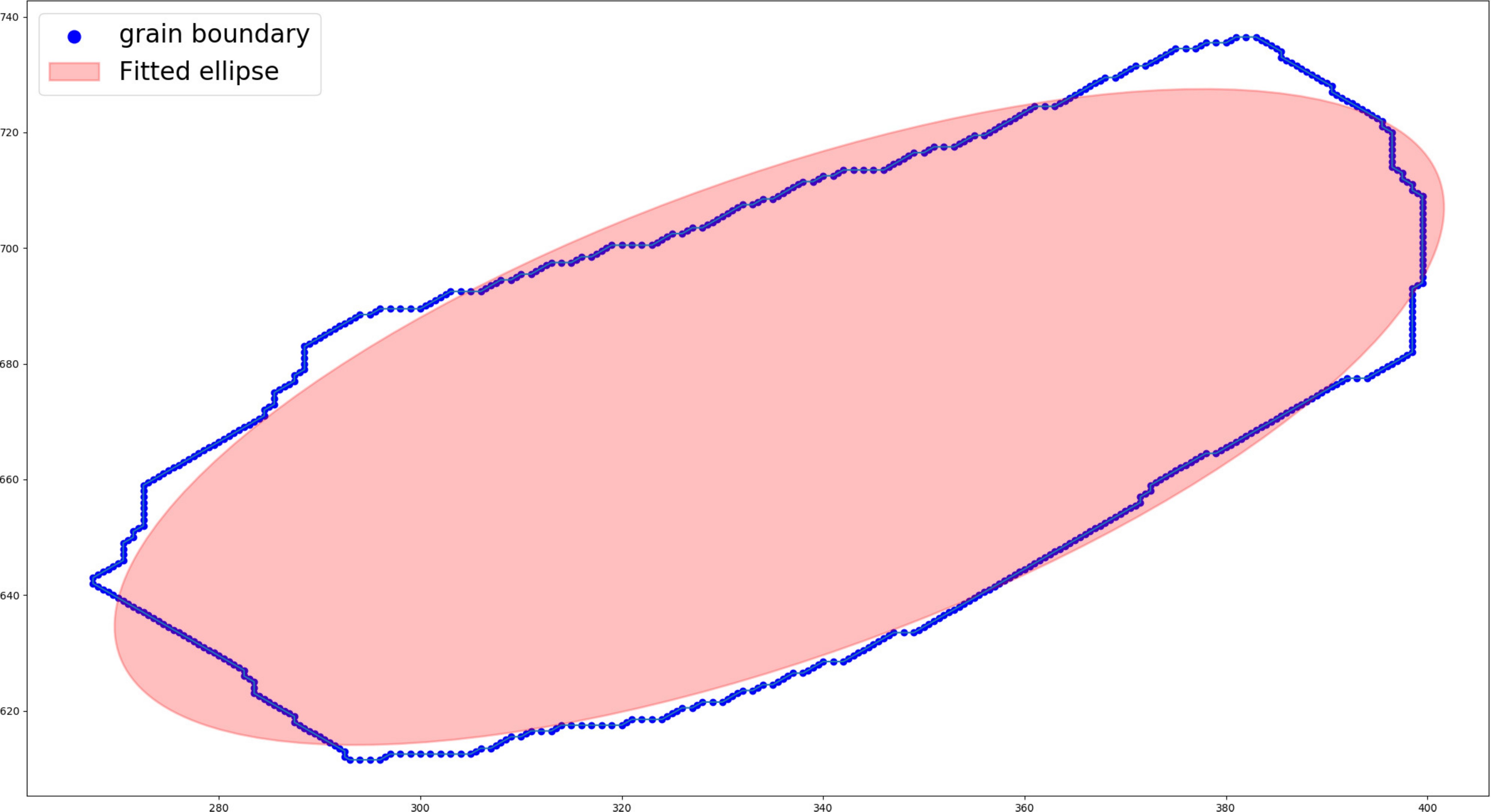}
	\caption{Best fit ellipse (rotated $90^{\circ}$ ccw) (Figure \ref{fig:cropped_uniqueGrain16166}).}
	\label{fig:cropped_fittedEllipse}
\end{subfigure}
	
\caption{Fitting ellipse process for a single grain in a microstructure. For each grain, the best fit ellipse is found by minimizing the total least square error. The distributions of the major dimension $a$, minor dimension $b$, locations of the center $(x_c, y_c)$, orientation of the ellipses $\theta$, are used to characterize the microstructure. There are also other microstructure descriptors.}
\label{fig:fitEllipseProcedure}
\end{figure}

The microstructure descriptors used are listed in Table \ref{tab:msDescriptors11}, which contains 11 microstructure descriptors. 
The microstructure descriptors associated with the best fit ellipse are described as follows. 
Each grain is approximated using a best fit 2D ellipse. 
The parameters describing the ellipse are then used as physics-based microstructure descriptors, including the dimensions of major and minor axes, the orientation, the coordinates of the best fit ellipse for the grain, as well as the grain area. The best fit ellipse is found by minimizing the total square error, which measures the difference between the grain boundary and the ellipse. The grain boundary is obtained by segmenting the grain from the microstructure. 
Figure \ref{fig:fitEllipseProcedure} describes the process of fitting an ellipse to the grain, and demonstrates it using a representative grain (Figure \ref{fig:cropped_uniqueGrain16166}) in the microstructure (Figure \ref{fig:microstructureExample}).


\begin{figure}[!htbp]
\includegraphics[width=1\textwidth,keepaspectratio]{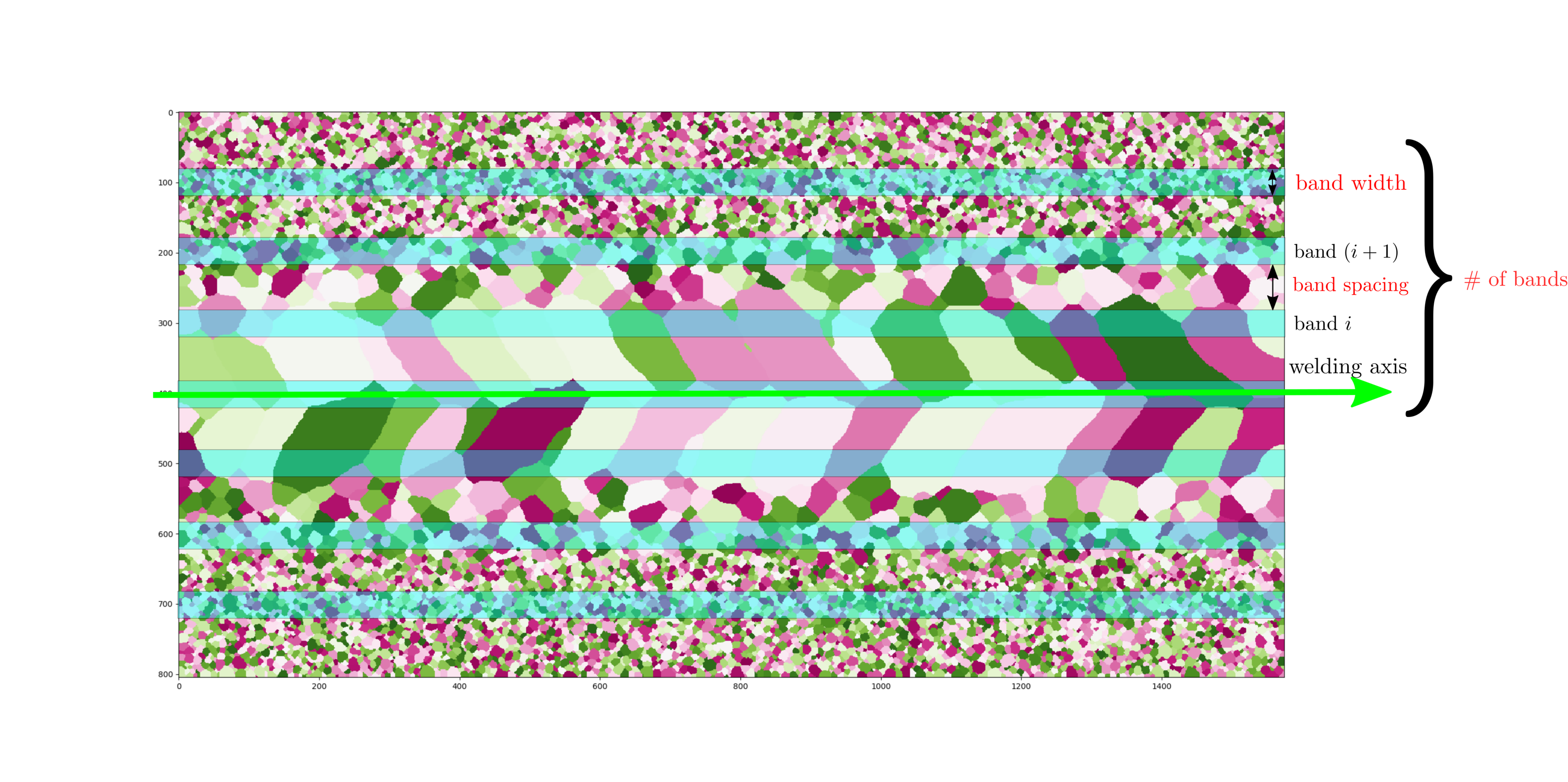}
\caption{Local sampling chord-length distribution along welding direction ($x$-direction) for the microstructure shown in Figure \ref{fig:microstructureExample}. Input parameters include the width of the bands, the spacing between bands, and the number of the bands. Distributions are assumed to be symmetric across the welding axis (denoted as green arrow).}
\label{fig:cropped_microstructureExample_ChordLengthIllustration}
\end{figure}

Since the microstructure descriptor choices are left to the users, it is possible to choose local microstructure descriptors in such a way so that the microstructures can be compared locally, e.g. by fixing at a particular location or angle. 
As a result of the local microstructure descriptors, the heterogeneous and anisotropic behaviors can be captured. 
Because the welding microstructure is translationally invariant along the welding axis, theoretically, the density of the chord-length does not change when it is sampled at any $x$-location, as in Figure \ref{fig:cropped_microstructureExample_ChordLengthIllustration}. 
However, the chord-length statistics change drastically, depending on the $y$-location. 
It is reasonable to assume symmetricity across the welding axis, but shifting up or down would affect the chord-length statistics. 
Thus, the $y$-location dependence allows one to capture the heterogeneity of the microstructure. 



In Figure \ref{fig:cropped_microstructureExample_ChordLengthIllustration}, the local chord-length statistics are obtained through the horizontal bands, which are imposed in parallel with the welding axis. 
The width of the band, the spacing between bands, and the number of bands are used to generate the bands. When one microstructure is compared with another microstructure, the chord-length distribution for each band is built and compared \textit{individually} with another. 
This approach allows one to compare microstructure locally, and thus also allows one to compare microstructure in a heterogeneous and anisotropic manner. 
For example, the directions and locations of the bands can change to capture the most distinguished microstructure features.

\subsection{Conditional statistical microstructure descriptors by imposing filters}

Before comparing the microstructure quantitatively, it is important to eliminate some of the microstructure attributed by the base metal. 
In general, the base metal contributes most of the grains, which are small, in the grain population. 
When comparing two microstructures, if the base metal grains are not excluded, then the two microstructures are very similar because their base metal grains are similar and the base metal grains is the majority of the microstructure. 
This effect is undesirable, as the base metal grains are considered as background and not the main focus when comparing two microstructures. 
Thus, it is necessary to exclude the base metal microstructure. 
Since the base metal grains are small, a grain area filter is used to exclude these grains. 

\begin{figure}
\centering
\includegraphics[width=0.75\textwidth,keepaspectratio]{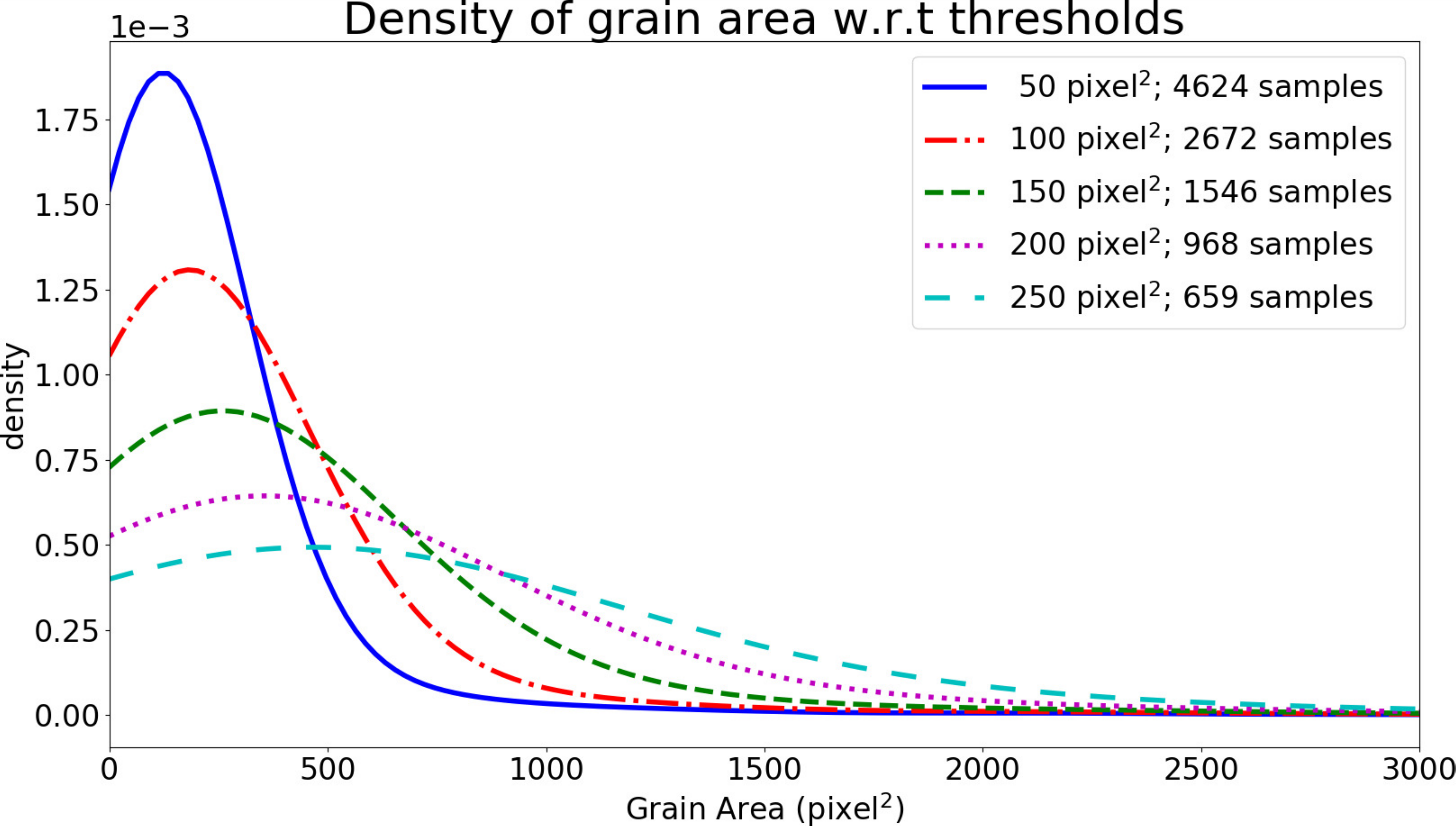}
\caption{Approximated conditional probability density function of grain area (a microstructure descriptor) with various thresholds on grain area to filter grains. If the threshold is large, the number of samples (grains) decreases, resulting in a higher approximation error. If the threshold is low, it is hard to quantitatively distinguish the candidate and target microstructures, due to the dominant portion of samples from the base metals.}
\label{fig:thresholdDiff}
\end{figure}

Figure \ref{fig:thresholdDiff} shows the effects of a low and high threshold in filtering out the grains on a representative microstructure before the probability density functions of the statistical microstructure descriptors are computed. 
The number of samples decreases monotonically, as 4624 grains, 2672 grains, 1546 grains, 968 grains, 659 grains, as the threshold on grain area increases from 50, 100, 150, 200, and 250 pixel$^2$. 
When filtering on grain area is disabled, the probability density function is heavily shifted to the left, because the grain population associated with the base metal region is dominant, compared to the grain population associated with the heat-affected zone and the melting zone.

It is straightforward to see that the difference metric\black{, i.e. the scalarized single-objective function,} and noise increase with increasing threshold. 
On one hand, if the threshold is too large, the statistical descriptors are noisy \black{because there might not be enough samples to approximate a probability density function}, thus resulting in noisy objectives, and therefore the multi-objective optimization problem becomes very challenging to solve due to the noise. 
On the other hand, if the threshold is too small, for example, no threshold, the statistical descriptors are close to each other, and the differences between objectives are almost at the same magnitude as the noise, and again, the optimization problem is too noisy to solve robustly. 
Thus, there is an optimal region of threshold that induces the highest signal-to-noise ratio. 
This optimal threshold is hard to find, but can be roughly estimated based on the preliminary examination of the microstructures. 
As the main goal is to eliminate the effect of base metal microstructure, the threshold can be slightly larger than the expectation of the physical descriptors of the base metal microstructure. 
The signal-to-noise ratio does not have to achieve a maximum number in order to utilize the proposed framework.

Therefore, it may be necessary to filter most of the base metal microstructure since the effects of scale separation are fairly severe. 
In this work, a grain area filter is implemented. 
If the grain area is larger than the threshold, then it is considered as a sample in the population. 
Otherwise, it is simply disregarded when microstructure descriptors are computed. 
The threshold of 150 pixel$^2$ is chosen by analyzing its effect on the studied microstructures. 

\subsection{Quantifying aleatory uncertainty of the objectives for the target microstructure}

\begin{table}[!htbp]
\caption{Quantified aleatory uncertainty of each single-objective that corresponds with the microstructure descriptors described in Table \ref{tab:msDescriptors11}.}
\label{tab:msDescriptorsNoise}
\centering
\scriptsize
\begin{tabular}{|c|l|l|} \hline 

Associated Obj. Fun. & Mean $\E(y_i)$ & Variance $\Var(y_i)$ \\ \hline 
$y_1$    & 0.0112 & 3.7129e-03 \\ 
$y_2$    & 0.0157 & 1.0870e-02 \\ 
$y_3$    & 0.0002 & 6.7116e-05 \\ 
$y_4$    & 0.0100 & 2.3328e-03 \\ 
$y_5$    & 0.0065 & 3.4961e-03 \\ 
$y_6$    & 0.0029 & 1.1344e-03 \\ 
$y_7$    & 0.0114 & 1.0709e-02 \\ 
$y_8$    & 0.0024 & 1.7863e-03 \\ 
$y_9$    & 0.0065 & 9.2228e-04 \\ 
$y_{10}$ & 0.0037 & 1.2436e-03 \\ 
$y_{11}$ & 0.0098 & 4.1146e-03 \\ \hline
 & $\sum_{i=1}^{s=11} \E(y_i) = 0.0809$ & $\sum_{i=1}^{s=11} \Var(y_i) =(0.0170)^2$ \\ \hline
\end{tabular}
\normalsize
\end{table}

To characterize the intrinsic noise of the kMC simulation, we sample the kMC simulation 25 times for the same set of input parameters corresponding to the target microstructure. 
For each time, a different seed in the pseudo-random number generator is chosen. 
We then compare the $y_i = \mathcal{S}_i \Big( p_{D_i} (d_i | \textbf{sampleTargetMs}), \  p_{D_i} (d_i | \textbf{targetMs}) \Big)$ in the same manner with Equation \ref{eq:KullbackLeiblerDivergence}; however, in this case, the sets of input parameters of \textbf{sampleTargetMs} and \textbf{targetMs} are identical. 
Table \ref{tab:msDescriptorsNoise} lists the mean (i.e., $\E[y_i]$) and variance (i.e. $\Var[y_i]$) for 11 objective functions $y_i$ across 25 samples. 
It is noteworthy to point out that the noise associated the objective functions is fairly small, resulting in an acceptably large signal-to-noise ratio, so that the noise does not have a significant impact on the optimization process. 
Summing all the means and variances of each individual microstructure descriptors, the mean and the variance of the single-objective $y = \sum_{i=1}^{11}$ are 0.0809 and $(0.0170)^2$, respectively.






\subsection{Noisy asynchronously-parallel multi-objective Bayesian optimization}
\label{subsec:moBO}

After the 11 probability density functions of microstructure descriptors are collected, the Kullback-Leibler divergence, described in Equation \ref{eq:KullbackLeiblerDivergence}, is utilized to measure the difference between the candidate and the target microstructures. 
To convert from a multi-objective optimization problem to a single-objective optimization problem, $s=11$ multi-objective functions $\{y_i\}_{i=1}^s$ are simply summed to form a single-objective $y=\sum_{i=1}^s y_i$. 
The parallel Bayesian optimization aphBO-2GP-3B \cite{tran2020aphbo} framework is utilized, with 20 concurrent kMC simulations utilized for the first batch of the framework, while 5 concurrent kMC simulations are used to explore the input space of the processing parameters. 
Because all the simulations are feasible and there are no hidden constraints, the size of the third batch is set as zero, since no classification effort is needed. 
Python scripts are devised to quantify the microstructure descriptors from SPPARKS dump files. 
Scipy toolbox \cite{scipy} is used to reconstruct the kernel density estimation. 
3965 samples are collected at the end of this study, which takes roughly about 7-8 days of simulations. 
The Bayesian optimization package is prototyped in MATLAB, and interfaces between Python, MATLAB, and SPPARKS are constructed so that the case study can run continuously.

\subsection{Optimal results}
\label{subsec:OptimalResultsCaseStudy1}

\begin{figure}[!htbp]
\includegraphics[width=\textwidth,keepaspectratio]{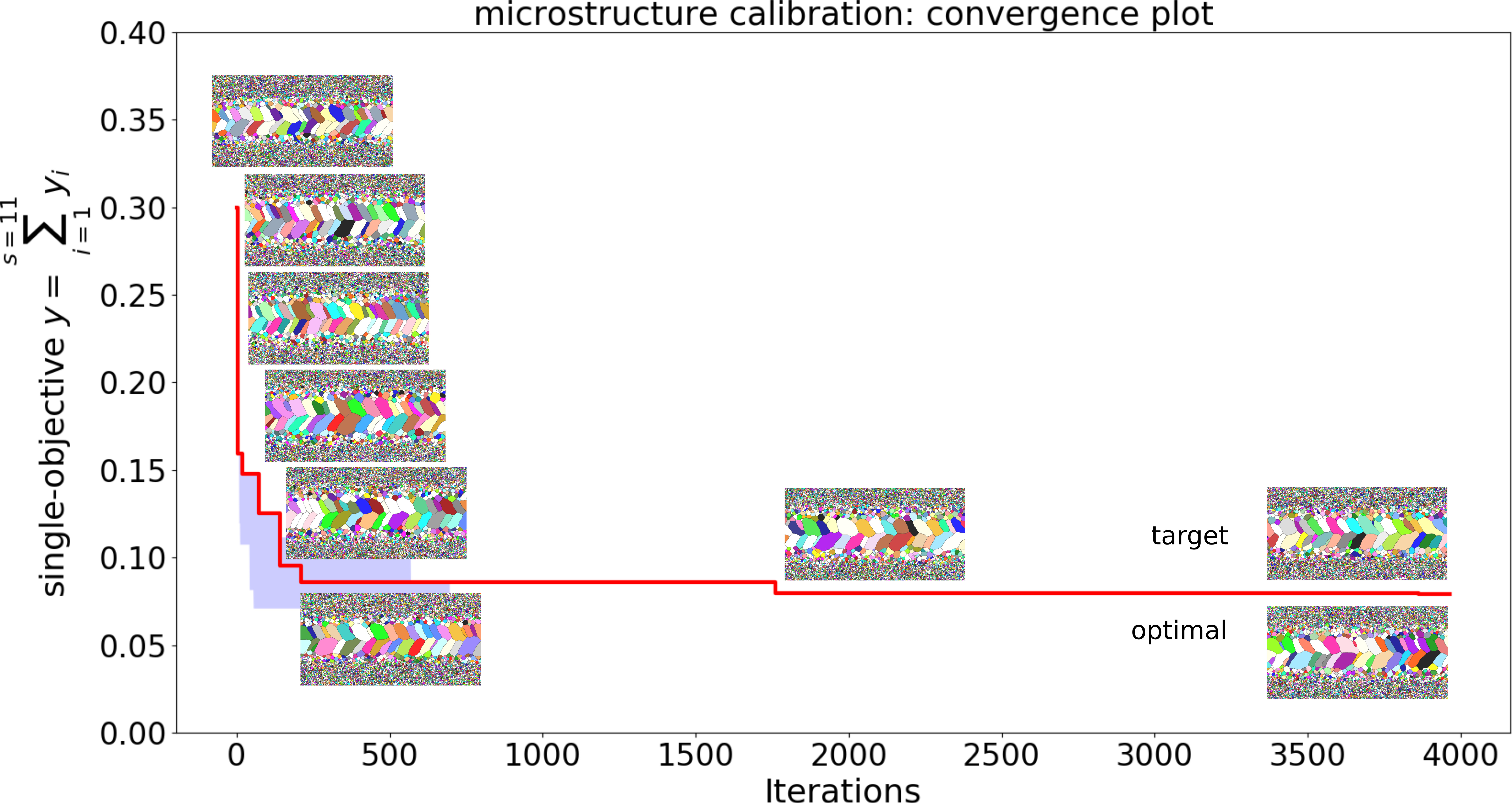}
\caption{Convergence plot of the microstructure calibration framework in kMC/welding problem. The objective is minimized as the optimization process advances. 
}
\label{fig:cropped_convergencePlotWithMsWeld2}
\end{figure}

Figure \ref{fig:cropped_convergencePlotWithMsWeld2} shows the convergence plot of the optimization problem, where the single-objective $y = \sum_{i=1}^{11} y_i$ is minimized as the optimization process advances. 
\black{ 
For the kMC welding case study, the following sampling points are used through Monte Carlo sampling to approximate the initial Gaussian process \black{model} response: (16.0, 135.0, 180.7), (18.0, 134.0, 165.2), (25.0, 145.0, 148.0), (29.0, 175.0, 155.2), (29.0, 169.0, 237.7), (19.0, 179.0, 103.50), (30.0, 200.0, 224.0). 
Three optimization runs are carried out to demonstrate the robustness of the microstructure calibration framework, where two out of these three runs last about 400 iterations for the purpose of demonstration, while the main one advances 4000 iterations. 
Here, one iteration corresponds to a single kMC simulation with different parameters. 
The converged objectives with respect to iterations are plotted in the blue envelope, representing the upper and lower bounds of objectives. 
In the following text, we will carefully analyze the main optimization run of 4000 iterations. 
}
Each simulation is performed on a computing cluster with the SLURM scheduler, where 36 cores are utilized on one node. 
On the average, the computational cost of one kMC simulation is around 9 CPU hours. 25 simulations are performed concurrently to reduce the physical waiting wall-time. The single-objective $y$ starts at the value of 0.29981499 at iteration 1, and subsequently converges to 0.15940848 at iteration 5, 
0.14780493 at iteration 21, 
0.12549670 at iteration 75, 
0.09523728 at iteration 142, 
0.08577485 at iteration 212, 
0.07945405 at iteration 1762, 
and 
0.07921463 at iteration 3864. 
Compared to the quantified aleatory uncertainty of the target microstructure in Table \ref{tab:msDescriptorsNoise}, $\E[y] = \sum_i \E[y_i] = 0.08086305$, the optimal objective $y = \sum_i y_i = 0.07921463$ is less than the quantified aleatory uncertainty, i.e. $0.07921463 < 0.08086305$. 
It is obvious to observe that the uncertainty of the single-objective $y$ has reached the quantified aleatory uncertainty; thus, it is unlikely to achieve better. 
Due to the low-dimensional problem, which contains only 3 parameterized inputs, the optimization has almost converged at iteration 212. The convergence of the optimization can be observed from the beginning to iteration 212. 
We conclude that the optimal microstructure is comparable to the target microstructure.


\begin{figure}
\begin{subfigure}[b]{0.475\textwidth}
	\centering
	\includegraphics[width=1\textwidth,keepaspectratio]{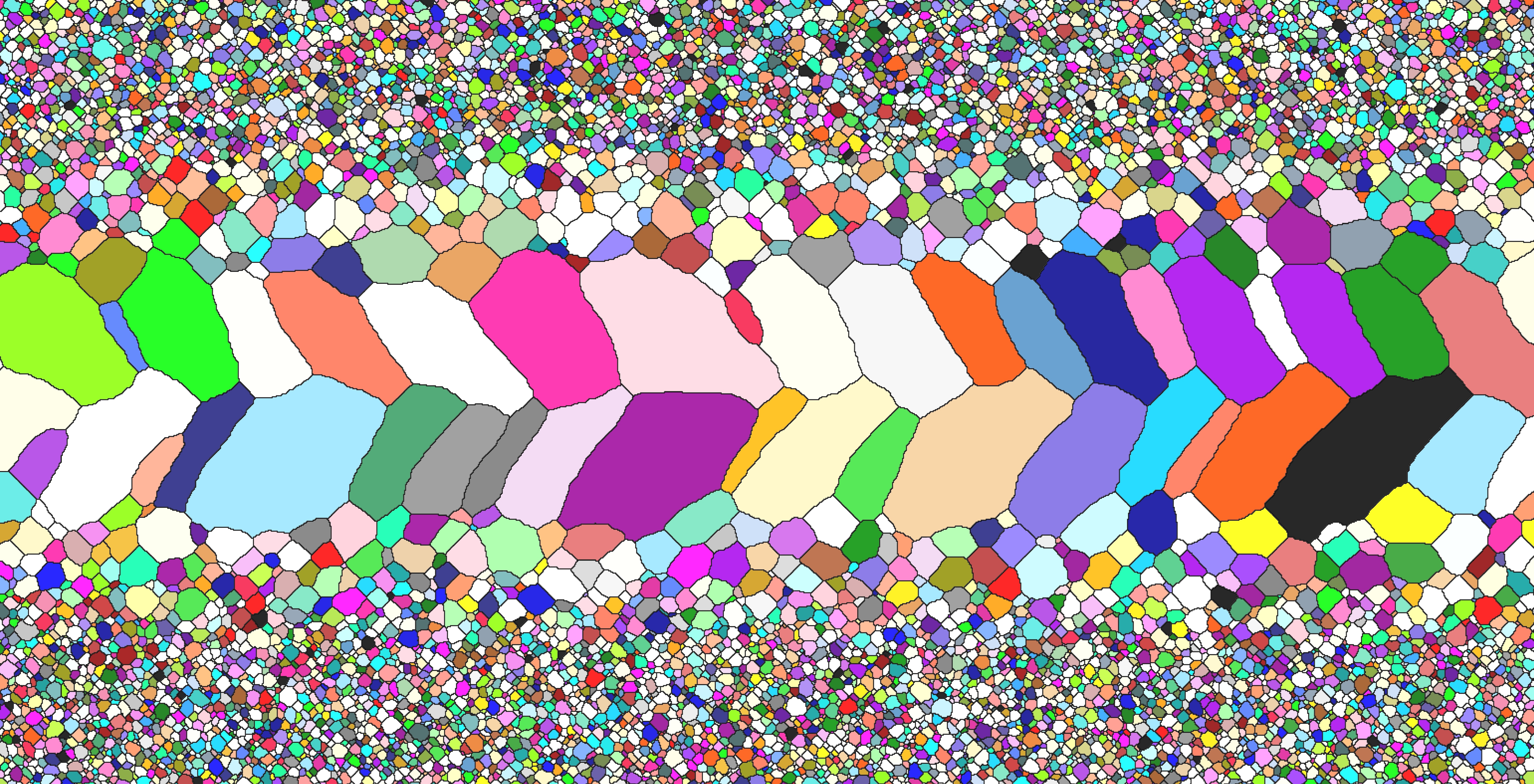}
	\caption{Optimal microstructure (Iteration 3864).}
	\label{fig:cropped_optimal_spkWeld_Iter3864}
\end{subfigure}
\hfill
\begin{subfigure}[b]{0.475\textwidth}
	\centering
	\includegraphics[width=1\textwidth,keepaspectratio]{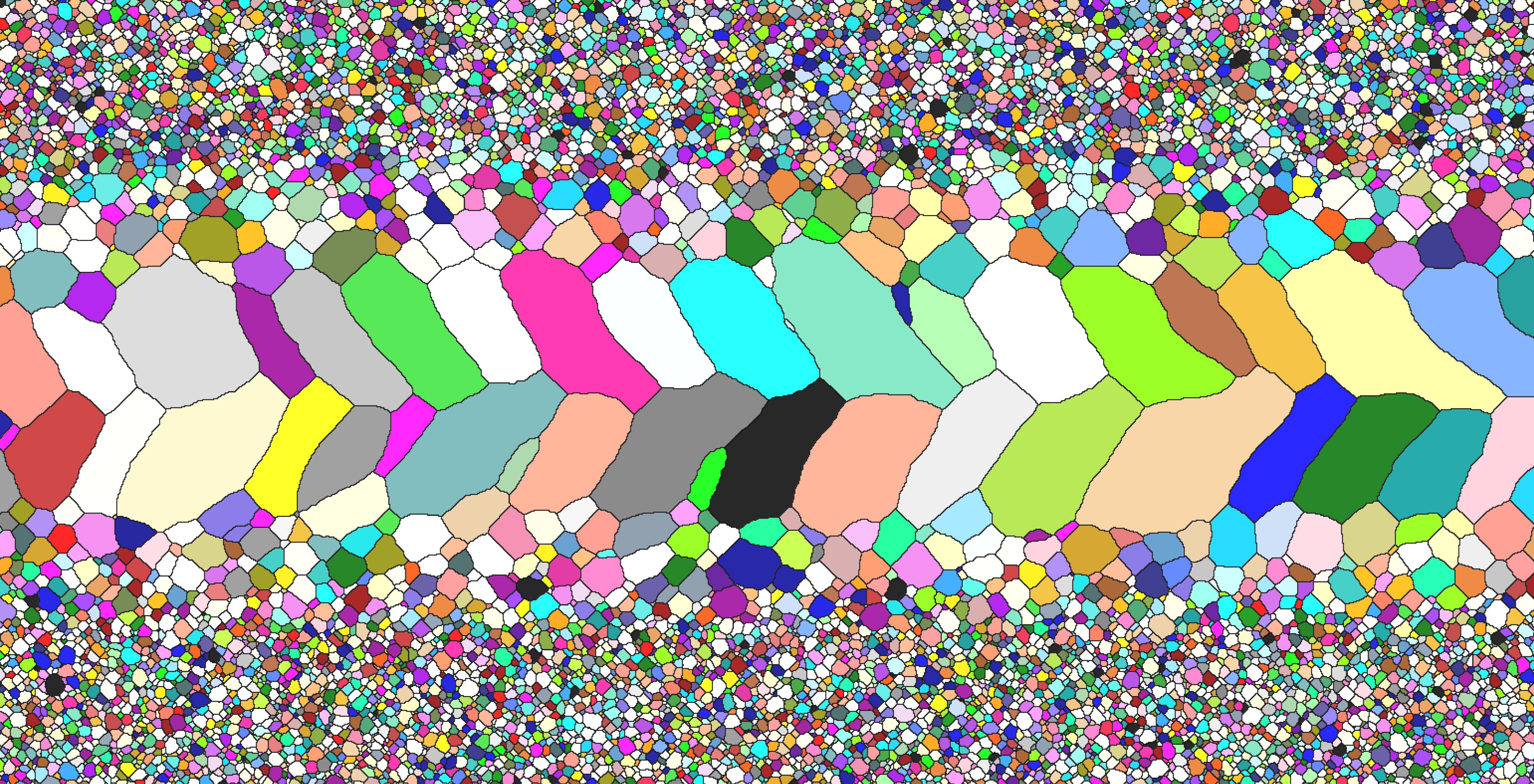}
	\caption{Target microstructure.}
	\label{fig:cropped_targetSample_spkWeld}
\end{subfigure}
\caption{Comparison between optimal and target microstructures.}
\label{fig:msComparison}
\end{figure}

Figure \ref{fig:cropped_optimal_spkWeld_Iter3864} and Figure \ref{fig:cropped_targetSample_spkWeld} show the optimal microstructure, obtained in iteration 3864, and the target microstructure, respectively. 
The inputs of the target microstructure are $(v,haz,width) = (15,150,200)$, as described in Table \ref{tab:inputTeardrop}, whereas the inputs of the optimal microstructure are $(v,haz,width) = (18.47182459, 154.86238483, 215.00287969)$. 
As observed in Figure \ref{fig:msComparison}, the dimension of the welding pool is accurately identified, as well as the velocity of the welding pool. 
Thus, we conclude that the microstructures in Figure \ref{fig:cropped_optimal_spkWeld_Iter3864} and in Figure \ref{fig:cropped_targetSample_spkWeld} are (almost) statistically equivalent. 

\begin{figure}
	\centering
	\includegraphics[width=\textwidth,keepaspectratio]{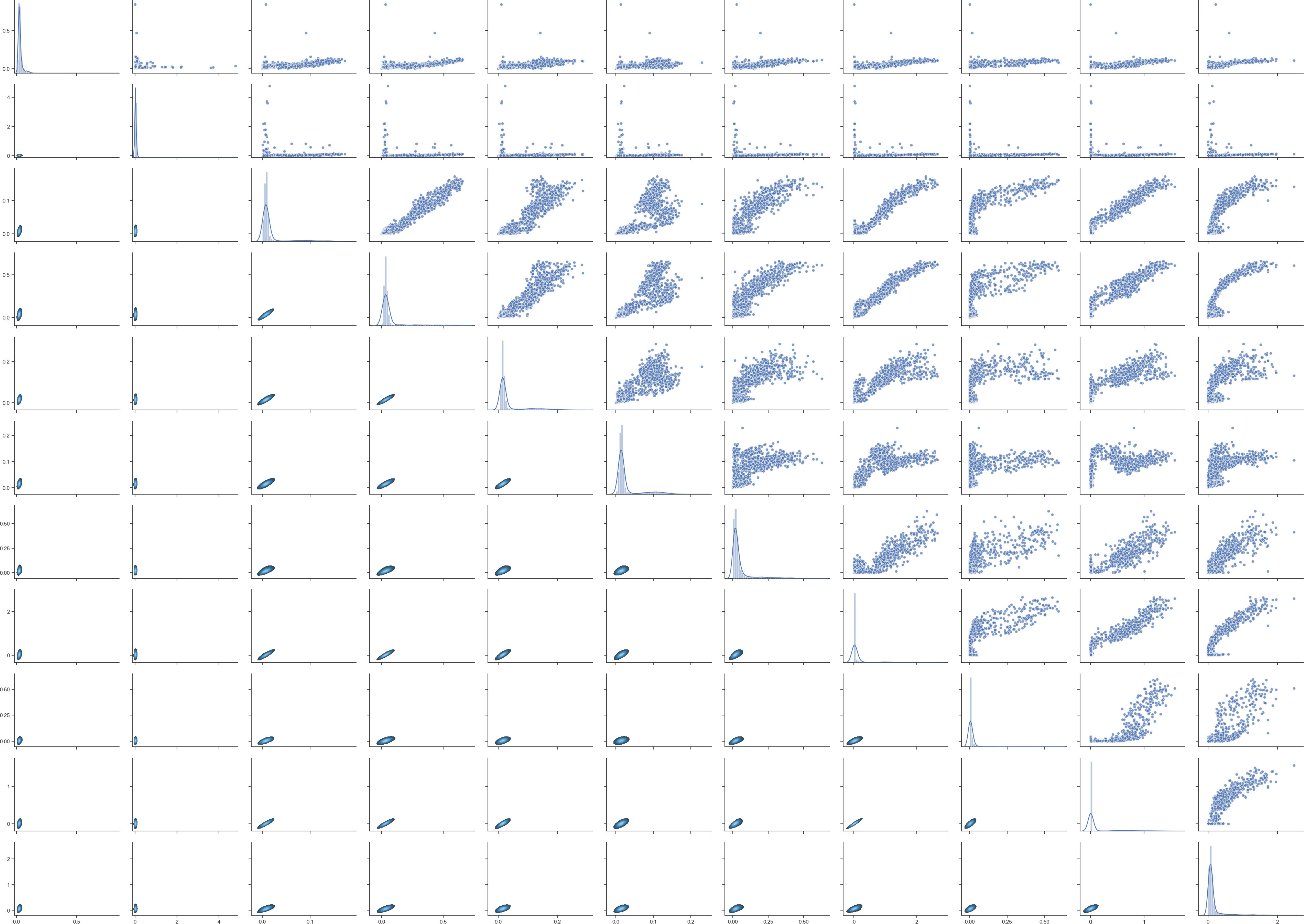}
	\caption{Scatterplot matrix between $11$ objectives $\{y_i\}_{i=1}^{s=11}$ (where the microstructure descriptor associated with each objective is listed in Table \ref{tab:msDescriptors11} with the same order) using 3965 samples collected at the end of the optimization process. 
	The lower part of the plot shows the joint density between $y_i$ and $y_j$ for $i \neq j$, and the upper part of the plot shows the scatter plot between $y_i$ and $y_j$. The diagonal of the plot shows the density of $y_i$. }
	\label{fig:cropped_scatterPlotMatrix}
\end{figure}

To further study the relationship between 11 objectives $\{y_i\}_{i=1}^{11}$, their correlations are investigated by reconstructing the joint density between $y_i$ and $y_j$. 
Figure \ref{fig:cropped_scatterPlotMatrix} shows the correlation between 11 objectives $y_i$, where the lower part of the plot shows the joint density between $y_i$ and $y_j$ for $i \neq j$, and the upper part of the plot shows the scatter plot between $y_i$ and $y_j$. The diagonal of the plot shows the density of $y_i$. 

\begin{figure}[!htbp]
\centering
\begin{subfigure}[b]{0.60\textwidth}
	\centering
	\includegraphics[width=0.75\textwidth,keepaspectratio]{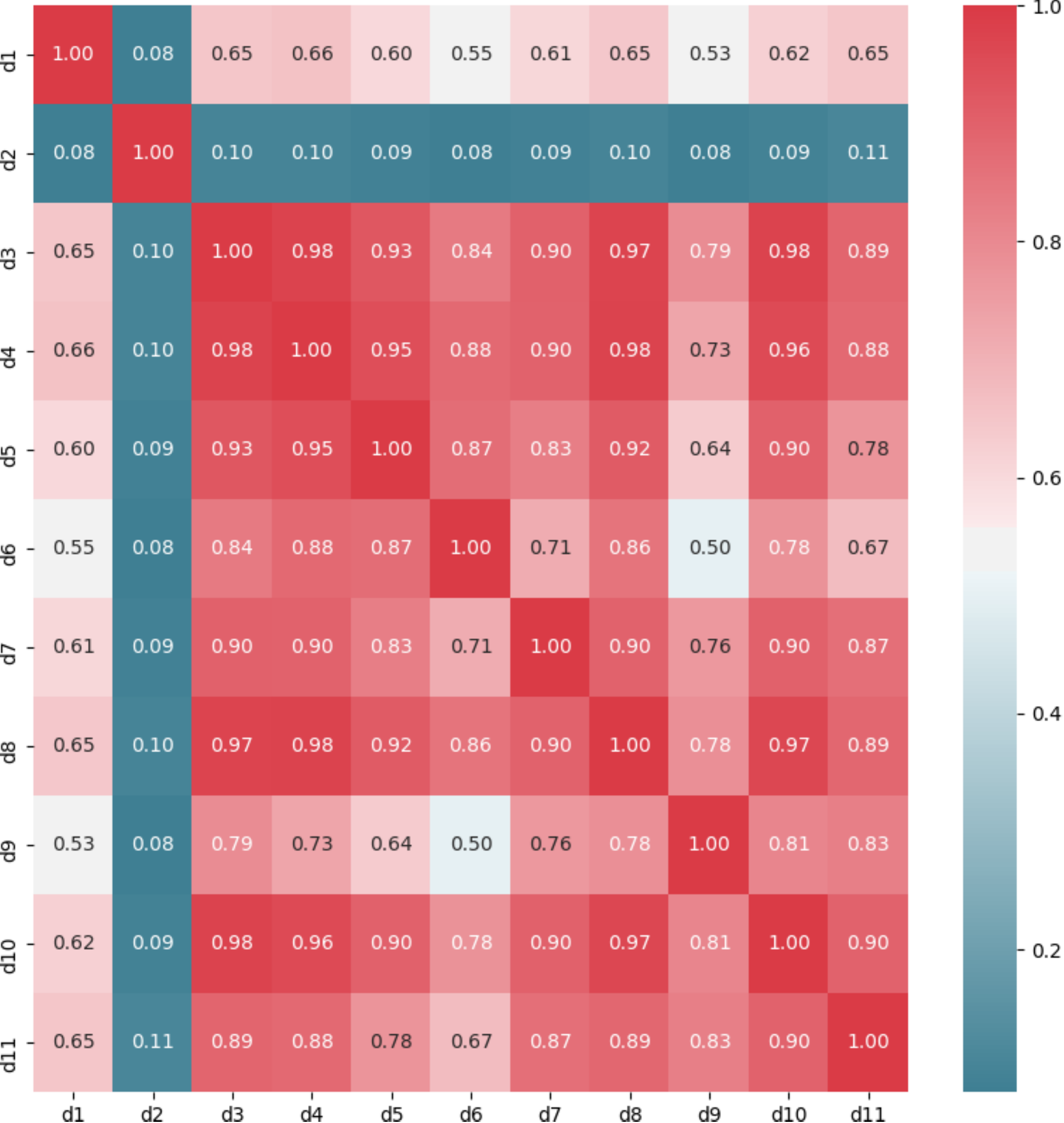}
	\caption{Diagonal correlation matrix between $11$ objectives $\{y_i\}_{i=1}^{s=11}$, where the coefficient of correlation $R^2$ is shown at the center of the square.}
	\label{fig:cropped_diagCorrMatrix}
\end{subfigure}
\vfill
\centering
\begin{subfigure}[b]{0.60\textwidth}
	\centering
	\includegraphics[width=1\textwidth,keepaspectratio]{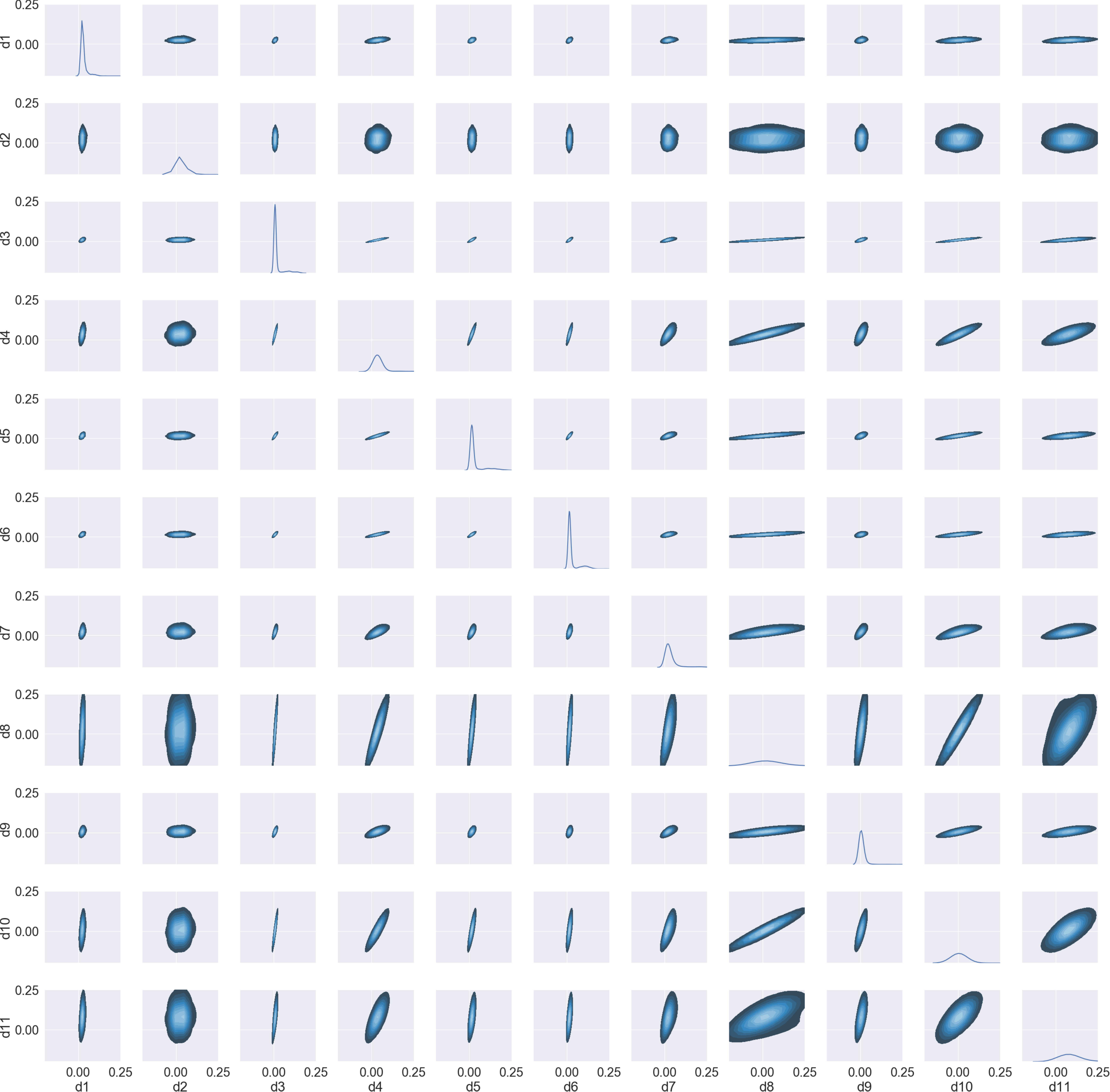}
	\caption{Magnified plot of the lower part in Figure \ref{fig:cropped_scatterPlotMatrix}, which shows the scatterplot matrix between objectives.}
	\label{fig:cropped_zoomKDEplot}
\end{subfigure}
\caption{Correlation matrix with the coefficient of correlation $R^2$ and the joint density function between 11 objectives listed in Table \ref{tab:msDescriptors11} with the same order. Most of them are strongly correlated with each others, while $y_2$ is nearly uncorrelated with the rest of the objectives.}

\end{figure}

Figure \ref{fig:cropped_diagCorrMatrix} shows the quantified correlation coefficient $R^2$ between two objectives. 
Figure \ref{fig:cropped_zoomKDEplot} shows the magnified plot of the lower part in Figure \ref{fig:cropped_scatterPlotMatrix}. 
Most of the microstructure descriptors are very strongly correlated, except for the $d_2 = b$ microstructure descriptor, which characterizes the dimension of minor axis of the best fit ellipse for the grain. 
In our experience, even though many microstructure descriptors are strongly correlated with others, including more descriptors does help in the optimization process. 

\section{Case study \#2: kMC simulation for grain growth and Bayesian optimization}
\label{sec:CaseStudy2}

\black{
In this section, we demonstrate the applicability of our \black{microstructure calibration framework} using another kMC simulation for grain growth problem, where a numerical temperature parameter is the input, and the microstructure is the output. 
The numerical temperature parameter is successfully recovered via our \black{microstructure calibration framework}. 
}

\subsection{Kinetic Monte Carlo simulation}

\black{
The details of temperature-dependent kMC simulation for grain growth and its implementation in SPPARKS is described in Garcia et al \cite{garcia2008three}, and is summarized here for the sake of completeness. 
In the grain growth simulation, the Potts model \cite{wu1982potts} is used to simulate curvature-driven grain growth. 
The Arrhenius equation describes the relationship between grain boundary mobility $M$ and temperature $T$ as 
\begin{equation}
M = M_0 \exp \left(\frac{-Q}{k_B T} \right),
\end{equation}
where $k_B$ is the Boltzmann constant, $M_0$ is the Arrhenius prefactor. 
The Metropolis algorithm is used to determine the probability $P$ of successful change in grain site orientation as 
\begin{equation}
P =
\begin{cases}
\exp\left( \frac{-\Delta E}{k_B T_s} \right), & \text{ if } \Delta E > 0, \\
1, &  \text{ if } \Delta E \leq 0, 
\end{cases}
\end{equation}
where $E$ is the total grain boundary energy calculated by summing all the neighbor interaction energies, $\Delta E$ can be regarded as the activation, and $T_s$ is the simulation temperature. 
It is worthy to note that the $T_s$ simulation temperature is not the real system temperature: $k_B T_s$ is an energy that defines the thermal fluctuation, i.e. noise, presented in the kMC simulation \cite{garcia2008three}. 
}

\subsection{Statistical microstructure descriptors as outputs}

\begin{figure}[!htbp]
\centering
\includegraphics[width=0.75\textwidth,keepaspectratio]{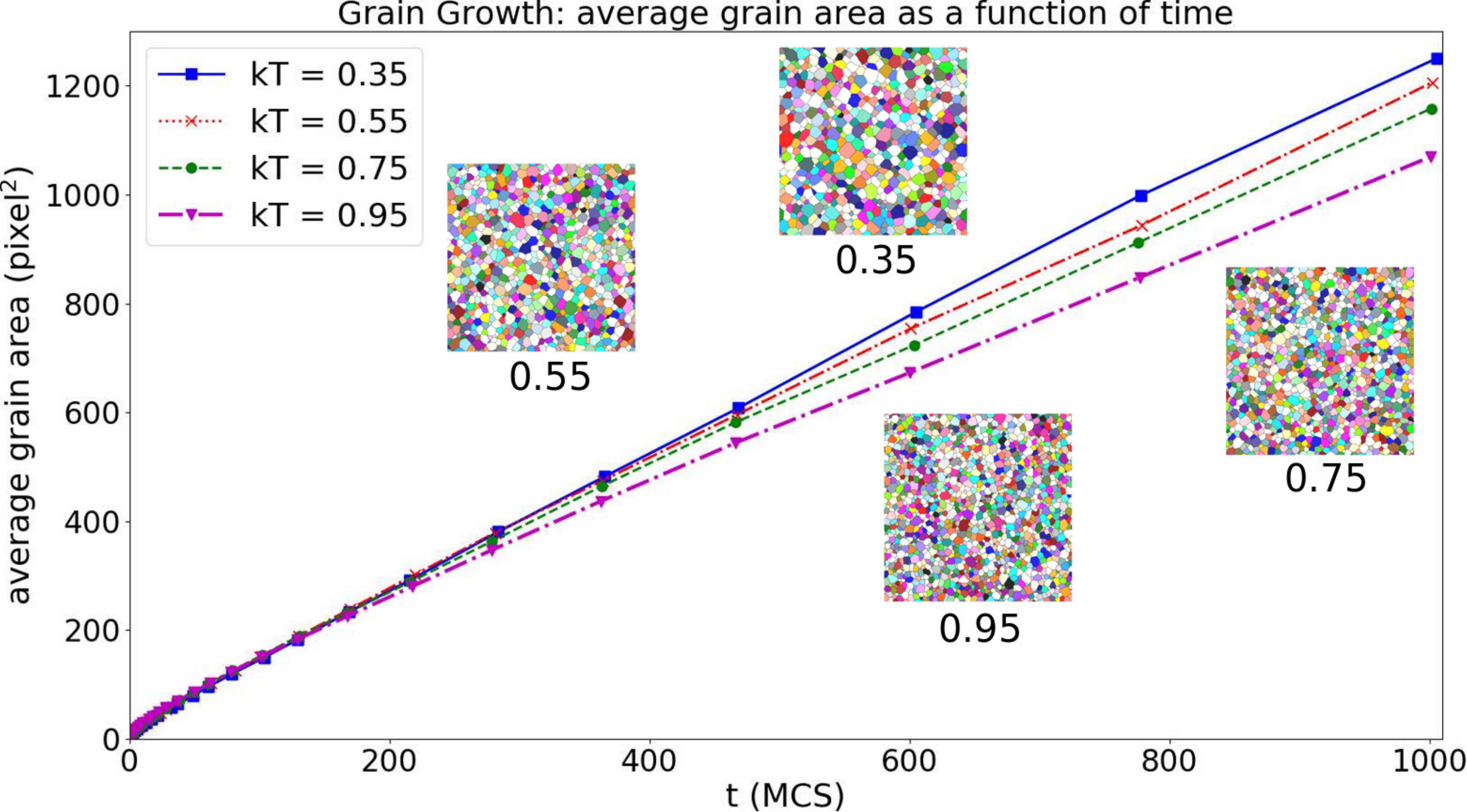}
\caption{Effect of the numerical temperature $k_B T_s$ in the kMC grain growth problem.}
\label{fig:cropped_gg-kMC-avgGrainArea}
\end{figure}

\black{
Figure \ref{fig:cropped_gg-kMC-avgGrainArea} shows the effect of $k_B T_s$ on different grain growth microstructure. In particular, larger $k_B T_s$ is associated with microstructures with smaller grain size for (nearly) the same amount of time. 
This suggests that the grain size distribution may be a good microstructure descriptor to describe and distinguish one microstructure from another. 
}

\black{
In this case study, the distribution of grain area ($p_{D_4}(d_4)$ in Table \ref{tab:msDescriptors11}) is used as the microstructure descriptor, where the Kullback-Leibler divergence measuring the distance between candidate and target microstructures is used as an output, where $k_B T_s$ is the numerical input describing simulation temperature. 
Because the number of grains are fairly small (few thousands), no filter is imposed on the statistical microstructure descriptors. 
In the target microstructure, $k_B T_s$ is set as 0.70, where the range of candidate microstructures is $[0.25, 0.95]$. 
The size of the simulation domain is 1024$\times$1024, which takes about 300s with 18 processors ($\approx 1.5$ CPU hours). 
}

\subsection{Optimal results}
\label{subsec:OptimalResultsCaseStudy2}



\begin{figure}[!htbp]
\includegraphics[width=\textwidth,keepaspectratio]{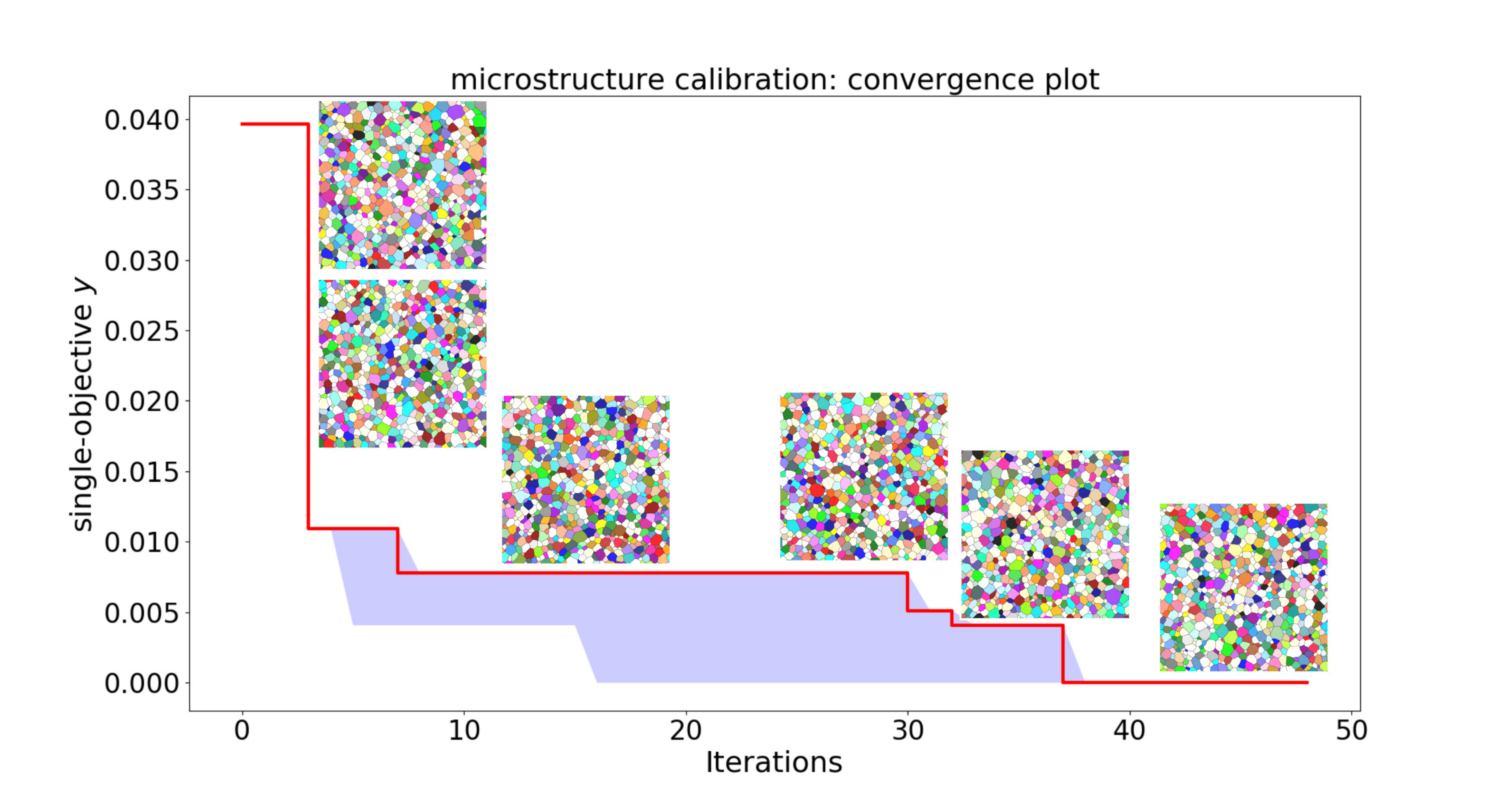}
\caption{Convergence plot of the microstructure calibration framework in kMC/grain growth problem. The objective is minimized as the optimization process advances. 
}
\label{fig:cropped_convergencePlotWithMsGG}
\end{figure}

\black{
Three optimization runs are carried out to demonstrate the robustness of the microstructure calibration framework for grain growth problem in kMC, where each last about 40-50 runs to obtain the correct target microstructure with $k_B T_s = 0.70$. 
The upper and lower bounds of objectives for three optimization runs are plotted as blue envelope in Figure \ref{fig:cropped_convergencePlotWithMsGG}. 
Three sampling points at $k_B T_s \in \{ 0.45,0.25,0.95\}$ are initialized. 
Four kMC simulations are performed concurrently across a high-performance computing platform. 
Among these three runs, we analyze thoroughly one run. 
Here, one iteration corresponds to a single kMC simulation with different parameters. 
The single-objective $y$ starts at the 
the value of 0.03965973 at iteration 1, and subsequently converges to 
the value of 0.01092325 at iteration 4, 
the value of 0.00776526 at iteration 8, 
the value of 0.00512348 at iteration 31, 
the value of 0.00408244 at iteration 33, and 
the value of 0.00000000 at iteration 38. 
The optimal input parameter $k_B T_s$ is 0.70393395, which agrees very well with the target input parameter $k_B T_s$ of 0.70. 
}

\section{Discussion}
\label{sec:Discussion}

\black{Microstructure descriptors are the key aspect of the microstructure calibration problem. }
First, the population of the microstructure descriptors is limited; for example, the number of samples is limited by the number of grains. 
Increasing the number of grains by increasing specimen size helps to reduce the approximation error in reconstructing the probability density function. 
Second, the microstructure in itself is also stochastic, where noise is due to an underlying aleatory uncertainty. 
\black{Depending on specific problems, the selection (i.e. inclusion or exclusion) of microstructure descriptors may positively or negatively affect the efficiency of the microstructure calibration framework. 
In our experience, including as many microstructure descriptors as possible may yield a better results, because some microstructure descriptors are capable of quantitatively distinguishing microstructures and some are not. 
Without any prior knowledge regarding the performance of the microstructure descriptors, it may be beneficial to include as many microstructure descriptors as possible. 
If the microstructure descriptor is not capable of distinguishing microstructure, it may be regarded as a noise in the objective function. 
Therefore, the microstructure calibration framework provides a mathematically sound and general approach to solve the inverse structure-process problems in the sense that it would always seek for the global optimal point, which corresponds to processing parameters of the target microstructure, given any predictive forward ICME model. 
}

As shown previously, it is desirable to quantitatively distinguish microstructures. 
In the best scenario, the microstructure descriptors should be uncorrelated, so that they can fully characterize the microstructure. 
However, in practice, this is typically not the case; for example, grain area should correlate at least fairly well with chord-length, since larger grains have longer chord-lengths and larger grain areas and vice versa. 
\black{
One of the advantages of the microstructure calibration framework is that microstructures can be compared quantitatively and not just qualitatively. 
In many cases, the human perception for qualitative comparison may be intuitively misleading and biased. 
Although the human eye may not be able to distinguish between these intermediate microstructures, they do have different microstructure features and are associated with different microstructure descriptors. 
Microstructures may be quantitatively different although they may look qualitatively similar. 
Quantitative comparison of microstructures through microstructure descriptors is therefore a robust measure to distinguish microstructures. 
}

A weighted formulation can be applied to convert multi-objective to single-objective problems. 
The weights can be adaptively computed based on cross-correlations, as well as the spatial gradients of microstructure descriptors. Choosing weights and norms, such as $L^2$, $L^{\infty}$, or $L^1$, to promote the difference between microstructures is a non-trivial research question and requires further study. 
In \black{the first case study}, the hyper-parameter \textit{bandWidth} (as shown in Figure \ref{fig:cropped_microstructureExample_ChordLengthIllustration}) used to locally capture the microstructure descriptor cannot be too large, because the homogeneous assumption is made across each band. 
This parameter should also not be too small, because it would lead to a small number of grains, thus resulting in a large approximation error in the approximated probability density function. 
The hyper-parameters for the microstructure descriptors to characterize the microstructure are chosen in an \textit{ad-hoc} manner. A more generalized framework for adaptively selecting microstructure descriptors may be helpful and more general. 



For additive manufacturing applications, if the tool path can be parameterized, a similar approach for microstructure calibration can be adopted to identify the tool path. 
The parameters generating the G-code, which defines the geometric model in additive manufacturing, can also be included as a part of the inputs, and the same microstructure calibration framework can be deployed to search for the target set of processing parameters, which includes the parameters to generate G-code and thus tool path. 

For 3D problems, a similar approach can be employed by generalizing the best fit ellipse algorithm in 2D to the best fit ellipsoid algorithm in 3D. 
Such approach requires a stable algorithm for fitting the ellipsoid in 3D and thus is posed as a future work. 
We also point out that 3D problems, in general, are much more complicated than 2D problems, in term of both computational complexity and microstructure descriptors. 
This topic is therefore posed as a potential future research. 

In this work, each microstructure descriptor is treated individually. 
However, a high-dimensional density function can also be adopted to the same formulation, by considering the joint probability density function of a set of microstructure descriptors. 
It is noteworthy that the approximation of the probability density function, in general, scales poorly with dimensionality. 
Further, the selection of microstructure descriptors is a non-trivial research question, which can be also considered in future work. 
For example, several groups of microstructure descriptors can be formed, where each group is associated with a high-dimensional probability density function, and subsequently an objective function in the multi-objective optimization framework. 
This interesting idea remains to be explored in future research. 

Since the microstructure is stochastic in nature, the microstructure calibration framework can be aligned with optimization under uncertainty. In particular, for a specific set of input parameters, multiple samples can be obtained to quantify aleatory uncertainty of microstructure descriptors. 
With experimental data, a multi-fidelity approach, such as CoKriging \cite{tran2019sbfbo2cogp,tran2020smfbo2cogp}, can be adopted to fuse predictions of computational and experimental data at the level of high-fidelity \cite{tran2019data}. 
For inputs with discrete and categorical variables, a mixed-integer Bayesian optimization approach \cite{tran2019constrained} can be utilized to search for the processing parameters, using the same formulation in this work. 
\black
{Multi-objective Bayesian optimization methods \cite{shu2020new,tran2020srmobo3gp} can be deployed to find the Pareto frontiers, which would provide computational insights for selecting objectives and weights for scalarization-based multi-objective Bayesian optimization approaches. }
Batch parallel Bayesian optimization methods, including both asynchronously parallel \cite{tran2020aphbo} and batch-sequential parallel \cite{tran2019pbo}, can be used to accelerate the optimization process. 
\black{
For some ICME applications where the intrinsic variance for microstructure descriptors is quite large and varies dependently with respect to the input parameters, more rigorous approaches, such as stochastic Gaussian process \black{model} (or stochastic kriging) \cite{ankenman2010stochastic,chen2017sequential} and composite Gaussian process \black{model} \cite{ba2012composite}, would be more suitable. 
}

It is noteworthy to point out that without parallelizing the Bayesian optimization by running 25 simulations concurrently, the sequential Bayesian optimization approach would still be at around iteration 200, while the parallel Bayesian optimization has already advanced to iteration 5000. 
For high-dimensional optimization problems, the improvement in convergence is more significant. 
The high-throughput feature in this framework is enabled by both parallelisms in optimization and in simulations.

The microstructure calibration framework, generally, is not restricted to the process-structure linkage, but also is highly applicable to model calibration applications, when model parameters are inferred based on the similarity of microstructures. 
Such problems are common in phase-field simulations, where thermodynamics parameters are often computed or fitted using some computational thermodynamic package \cite{xiong2010phase}. The microstructure calibration framework provides an alternative and viable solution to such model calibration problems, where microstructures are considered as objective functions.

\section{Conclusion}
\label{sec:Conclusion}

In this work, a microstructure calibration framework is proposed to efficiently solve the inverse problem in process-structure linkage. 
A set of microstructure descriptors is used to characterize and quantitatively compare microstructures. 
By measuring differences of microstructure descriptors densities, the inverse problem is formulated as a noisy multi-objective optimization problem. 
To that end, a parallel Bayesian optimization framework is utilized to minimize differences. 
As a result, the optimal microstructure is obtained and compared with the target microstructure. 
\black{Two case studies are demonstrated, where kMC simulations are employed in forward prediction. 
In both cases, the input parameters associated with the target microstructure are successfully recovered, thus demonstrating the usefulness of the proposed framework.}

\section*{Acknowledgment}

A.T. thanks Theron Rodgers (SNL) for numerous helpful discussions. 
\black{The authors are grateful to the anonymous reviewer for his or her constructive comments. }
The views expressed in the article do not necessarily represent the views of the U.S. Department of Energy or the United States Government. 
Sandia National Laboratories is a multimission laboratory managed and operated by National Technology and Engineering Solutions of Sandia, LLC., a wholly owned subsidiary of Honeywell International, Inc., for the U.S. Department of Energy's National Nuclear Security Administration under contract DE-NA-0003525. 
This research was supported by the U.S. Department of Energy, Office of Science, Early Career Research Program, under award 17020246, and by Sandia LDRD program 219144.

\bibliographystyle{elsarticle-num}
\bibliography{lib}

\end{document}